\documentclass{aastex6}
\usepackage[inline]{enumitem}

\newcommand{\kes}{\kappa_{\rm es}}
\newcommand{\kf}{\kappa'_{\rm *f}}
\newcommand{\ssb}{\sigma_{\rm sb}}

\newcommand{\music}[3][,]{\item \href{#2}{#3}{#1}}
\newenvironment{soundtrack}
    {\vskip6pt{\large\it Soundtrack:} \begin{enumerate*}[label={}]
    }
    {\end{enumerate*}
    }

\begin{document}

\title{Non-blackbody Disks Can Help Explain Inferred AGN Accretion Disk Sizes}

\author{Patrick B. Hall, Ghassan T. Sarrouh}
\affil{Department of Physics and Astronomy, York University, Toronto, ON M3J 1P3, Canada}
\author{Keith Horne}
\affil{SUPA Physics and Astronomy, University of St.\ Andrews, North Haugh, St.\ Andrews KY16 9SS, Scotland, UK}

\begin{abstract}
If the atmospheric density $\rho_{\rm atm}$
in the accretion disk of an active galactic nucleus (AGN) is sufficiently low, 
scattering in the atmosphere can produce a non-blackbody emergent spectrum.
For a given bolometric luminosity, at ultraviolet and optical wavelengths
such disks have lower fluxes and apparently larger sizes
as compared to disks that emit as blackbodies.
We show that models in which $\rho_{\rm atm}$ is a sufficiently low fixed fraction of 
the interior density $\rho$
can match the AGN STORM observations of NGC 5548
but produce disk spectral energy distributions 
that peak at shorter wavelengths than observed in luminous AGN in general.
Thus, scattering atmospheres can contribute to the explanation for large
inferred AGN accretion disk sizes but are unlikely to be the only contributor.
In the appendix section, we present unified equations for the interior $\rho$ and $T$ in 
gas pressure-dominated regions of a thin accretion disk. 
\end{abstract}


\section{Introduction} \label{sec:intro}

The AGN STORM project has measured the
wavelength-dependent time delays with which the accretion disk of
NGC 5548 reverberates in response to
emission from very close to the black hole \citep{2015ApJ...806..129E}.
The delays scale approximately
as $t_{\rm delay}(\lambda)\propto\lambda$ \citep[][hereafter F16]{2016ApJ...821...56F}.
This result has been interpreted as evidence for a disk that emits
as a blackbody with a surface temperature profile 
$T_S(r)\propto r^{-1}$ \citep{2017ApJ...835...65S},
since $t_{\rm delay}\propto r/c$ and 
the peak wavelength of a blackbody spectrum follows $\lambda\propto T^{-1}$. 
In contrast, a blackbody disk that locally releases
the energy generated by viscous dissipation has a slower
decrease of $T_S$ toward larger radii, $T_S(r)\propto r^{-3/4}$
(\citealt[][SS73]{ss73}; \citealt[][NT73]{1973blho.conf..343N}). 
A blackbody disk thus has
$t_{\rm delay}(\lambda)\propto\lambda^{4/3}$;
compared to NGC 5548, its time delays
would drop more quickly toward shorter wavelengths.

Furthermore, the lags in NGC 5548 imply a disk $\simeq$3 times larger
than expected for a standard thin disk with $L/L_{\rm Edd} \sim 0.1$ (F16).
AGN accretion disks with radii $2\times$ or more larger than expected
were first
inferred from microlensing observations of quasars \citep[e.g.,][]{2008A&A...490..933E,2010ApJ...712.1129M,2011ApJ...729...34B,2015ApJ...798...95B}.
Larger than expected sizes have also been deduced 
from photometric variability of quasars in Pan-STARRS \citep{JiangYF2017}
and DES \citep{DESadsizes}.
Proposed explanations for this discrepancy \citep[see \S4 of][]{ad} include 
an inhomogeneous disk with a range of temperatures at each radius 
(\citealt{da11}; see also \citealt{Cai+18EUCLIA}),
a long timescale for the disk to respond to changing illumination \citep{gd16},
and an increase in the mean measured lag due to reverberation from diffuse gas at
large scales \citep{Cackett+18NGC4593}.

Here, we explore another mechanism that can help explain the above observations:
a disk that deviates from a blackbody 
such that as the radius decreases, $\lambda_{\rm peak}$ shifts
to an increasingly shorter wavelength 
than for a blackbody of the same total flux.
Electron scattering in the disk atmosphere
of photons generated by free-free emission within the disk
can produce an emergent spectrum with just such a shift (\S \ref{twomodels}).
As originally discussed in NT73 \S 5.10 and SS73 \S 3a,
free-free emission produces shorter-wavelength photons deeper in the disk.
If scattering is non-negligible in the disk atmosphere,
such photons 
are less likely to escape;
the emergent spectrum becomes non-blackbody 
(e.g., \citealt{ce87}). 
To locally radiate all energy generated by viscous dissipation,
the disk must heat up: as scattering becomes sufficiently dominant
(e.g., at low atmospheric densities),
the peak of the spectrum shifts to shorter wavelengths
and disks appear larger at a given wavelength than in the blackbody case
(e.g., \citealt{1991ApJ...381L..39R}).
A small spectral shift in this direction (of insufficient magnitude to match observations)
can occur in the emission from disk regions where electron scattering dominates 
(\S 5.10 of NT73 and \S 3a of SS73).
A further shift in this direction (still of insufficient magnitude to match observations)
has been seen in previous work on disks with 
significant magnetic pressure support and thus low-density atmospheres
\citep[e.g.,][]{2006ApJ...645.1402B}.
This proposed explanation for large inferred accretion disk sizes
is related to that of \cite{birdsnest}, 
but distinct in that the electron scattering occurs in an accreting
inflow in that work and in the disk atmosphere in this model.
In \S \ref{ngc5548}, we explore which model parameters can
produce a shift of the magnitude required to match the observations of NGC 5548,
and we discuss our results in \S \ref{theend}.

\section{Two non-blackbody spectral models} \label{twomodels}

To examine how the spectrum from a disk with a scattering atmosphere varies with radius when gas pressure dominates,
we consider two atmospheric density distributions: constant and exponential.
In both cases, we assume the atmospheres to be isothermal with height.

\subsection{Constant-density Atmosphere} \label{conair}

For the form of the emergent specific flux $F_\nu(r)$, 
we adopt the modified blackbody spectrum appropriate for isotropic scattering in
a homogeneous medium (\citealt{RybickiLightman}; Equation 7 of \citealt{kph04}):
\begin{equation}
F_\nu(r) = \frac{2\pi B_\nu(T_S(r))}{1+\sqrt{1+\kappa_{\rm es}/\kappa_{\rm\nu,abs}(r)}}
\end{equation}
where we use $T_S$ to denote a characteristic disk surface temperature
and where $\kappa_{\rm es}$ and $\kappa_{\rm\nu,abs}$
are the electron-scattering opacity and the frequency-dependent absorption opacity, respectively.
The factor of $\pi$ comes from integrating the
(assumed isotropic) emergent specific intensity over all solid angles.

The value of $T_S(r)$ is found by requiring local energy release.
At each radius, the flux emitted over all wavelengths must equal 
the rate at which energy is generated per unit surface area:
$\int F_\nu(r) ~d\nu = D(r)$ where $D(r)$ is given by Eq.\ \ref{eqd}.
For $\kappa_{\rm\nu,abs} \gg \kappa_{\rm es}$, 
which occurs at sufficiently high densities or sufficiently low temperatures,
$F_\nu = \pi B_\nu(T_S)$ and $T_S(r)=(D(r)/\ssb)^{1/4}$.
For $\kappa_{\rm\nu,abs}\rightarrow 0$,
which occurs at sufficiently low densities or sufficiently high temperatures,
$F_\nu \rightarrow 2\pi B_\nu(T_S) \sqrt{\kappa_{\rm\nu,abs}/\kappa_{\rm es}}$,
which can deviate strongly from a blackbody spectrum.

\subsection{Exponential Atmosphere} \label{expair}

\citet[ZS69]{1969SvA....13..175Z} derive the expression for the specific flux emerging 
from a semi-infinite isothermal scattering atmosphere with an exponential density distribution $ \rho(z) = \rho_1 e^{-(z-z_1)/h_1} $ where 
$h_1$ is the atmospheric scale height
and $z_1$ is the height, above the disk plane, of the {\sl edge} of the atmosphere.
We take $h_1 = r^3 k_B T_S / GM\mu_p h$ where 
$ h=(k_B T r^3/GM\mu_p)^{1/2} $
is the disk scale height (see the text following Eq.\ 2.28 in SS73).
Thus, we have 
\begin{equation}
h_1 = \left(\frac{k_B T_S^2 r^3}{GM\mu_p T}\right)^{1/2} = h\frac{T_S}{T}.
\label{eq_h1}
\end{equation}
We define 
\begin{equation}
k_{\nu,1}\equiv 3\kes\kappa_{\rm\nu,abs1}h_1^2/\rho_1
\end{equation}
which is independent of density because
$\kappa_{\rm\nu,abs1}$, 
the absorption opacity at the edge of the atmosphere,
is proportional to the local density $\rho_1$ (see \S\ref{yo}).
We can then write the emergent specific flux from this exponential atmosphere as
\begin{equation}
E_\nu(r) = \frac{ \pi B_\nu(T_S(r)) }{1+3\kes h_1 \aleph_1/2k_{\nu,1}^{1/3} }
\end{equation}
(Eq.\ A1.9 of ZS69, 
divided by 2 so that $E_\nu\rightarrow\pi B_\nu$ when electron scattering is unimportant; see \citealt{RybickiLightman}).
In that equation, we define
\begin{equation}
\aleph_1 \equiv 
\frac{ K_{1/3}(\frac{2}{3}k_{\nu,1}^{1/2}\rho_1^{3/2} ) }{ k^{1/6}_{\nu,1}\rho_1^{1/2} K_{2/3}(\frac{2}{3}k_{\nu,1}^{1/2}\rho_1^{3/2}) } > 0
\end{equation}
where the $K_a()$ are modified Bessel functions of the second kind.
When $\rho_1$ is small the $\aleph_1$ term can dominate the denominator of $E_\nu$
and yield a spectrum differing greatly from a blackbody spectrum.

In the limit $k_{\nu,1}^{1/2}\rho_1^{3/2} \rightarrow \infty$,
$\aleph_1\rightarrow 0$ and $E_\nu\rightarrow \pi B_\nu$.
In the limit $\rho_1=0$, $\aleph_1 \simeq 70/51$ to $<$1\% accuracy
and we can use the definition of $k_{\nu,1}$ to write the limiting form of $E_\nu$ as
\begin{equation} 
E_\nu \simeq \pi B_\nu \frac{102k_{\nu,1}^{1/3}}{210\kes h_1 }
\simeq \pi B_\nu \frac{17k_{\nu,1}^{1/3}}{35\kes h_1 } \simeq 
\left(\frac{512\kes\kappa_{\rm\nu,abs1}h_1^2}{35\rho_1\kes^3 h_1^3}\right)^{1/3} \pi B_\nu 
\simeq 
\left(\frac{51\kappa_{\rm\nu,abs1}}{35\rho_1\kes^2 h_1}\right)^{1/3} \pi B_\nu(T_S(r)) 
\end{equation}
which as expected is a factor of $\simeq 2^{1/3}$ lower than given at
the end of Appendix 1 of ZS69.

\subsection{Parameter Dependences} \label{yo}

The non-blackbody frequency dependence of the spectrum emerging from the above
atmospheres arises from the frequency dependence of the absorption opacity.
We assume $\kappa_{\rm\nu,abs}$ 
is given by the sum of free-free and bound-free absorption opacities,
and we approximate them as having the same frequency dependence
(\citealt{meier}, \S 9.3.6.2). 
Defining $x\equiv h\nu/kT_S$, the specific absorption opacity can then be written 
\begin{equation}
\kappa_{\rm\nu,abs} = \kappa_{\rm bf+ff} = \kf \rho T_S^{-7/2} x^{-3} (1-e^{-x})
\end{equation}
with
\begin{equation}
\kf=(3.68\times 10^{22} {\rm ~cm^2~g^{-1}~(g~cm^{-3})^{-1}~K^{7/2}})[X+Y+1180Zf(T)](1+X)
\end{equation}
where the H, He, and metal mass fractions are
$X=0.71$, $Y=0.27$, and $Z=0.02$, respectively, for solar metallicity
(\citealt{meier}, \S 9.3), 
and $f(T)$ is the fraction of metals that are not ionized
(1 for neutral gas and 0 for fully ionized gas).
The free-free opacity is dominated by electrons from H and He,
while the bound-free opacity is dominated by electrons bound to metals.
We take $f(T)=0.5$, which yields 
$\kf=8.04\times 10^{23}$ cm$^2$ g$^{-1}$ with the above mass fractions.

We also adopt $\kes = 0.2(1+X) {\rm ~cm^2~g^{-1}}$
(\citealt{meier}, \S 9.3.6.1); 
more metal-rich gas with a lower $X$ has fewer electrons per gram and therefore a lower electron-scattering opacity.

We can now write out the full parameter dependencies of the specific flux $F_\nu(r)$ emergent from a constant-density atmosphere, again using $x\equiv h\nu/kT_S$ for convenience:
\begin{equation}
F_\nu(r) = \frac{ (4k_B^{3}/c^{2}h^{2})  [x^{3}/(e^{x}-1)]  T_S^{3}}{1+\sqrt{1+0.2x^3T_S^{7/2}/\kf \rho(1-e^{-x}) [X+Y+1180Zf(T)] }}
\end{equation}
where the $(1+X)$ dependencies in $\kes$ and $\kf$ have canceled out in the denominator.

The parameter dependencies of the specific flux $E_\nu(r)$ emergent from an exponential atmosphere can be written explicitly only in the low-density limit.  That limit may not always apply, so we do not give the explicit form here.

In calculating $\rho$ and $T$ as part of determining $E_\nu(r)$ or $F_\nu(r)$,
we adopt a mean mass per particle of 
\begin{equation}
\mu_p = \frac{m_p}{2X + 0.75Y + Z[0.56-0.5f(T)]}
\end{equation}
(\citealt{meier}, \S 9.3.1.1) 
where $m_p$ is the proton mass.
That expression is appropriate for gas that is completely ionized in H and He but in which metals range from neutral 
to fully ionized. 
For the parameters given above, $\mu_p = 0.614 m_p$.

\subsection{Calculating $T_S$, the emergent spectrum, and mean measured radii}

The emergent $F_\nu$ from a disk atmosphere depends on the opacity and thus the density {\em in the atmosphere}, which will be lower than the average interior density.
We adopt an atmospheric density $\rho_{\rm atm}(r)$ which
is a constant fraction of the interior density $\rho(r)$.
To find $F_\nu(r)$ or $E_\nu(r)$ thus requires knowing the gas density $\rho(r)$,
which, for gas pressure-dominated regions of an AGN accretion disk,
requires knowing the disk interior temperature $T(r)$.  In such regions,
we use the analytic expressions for $T(r)$ and $\rho(r)$ derived in the Appendix.
For radiation pressure-dominated disk regions, 
we calculate $\rho(r)$ and $T(r)$ using Eq.\ 2.11 and 2.12 of SS73, 
respectively.\footnote{There is a typo in Eq.\ 2.12 of SS73, which reads
$r^{-3/4}$ instead of the correct value of $r^{-3/8}$; see Eq.\ 5.9.10 of NT73.}
The transition between these two regions is the radius at which they yield equal values 
of $\rho(r)$.  We adopt radiation pressure-dominated values of $\rho(r)$ and $T(r)$ 
interior to that radius, and gas pressure-dominated values exterior to that radius.

Once a value for the atmospheric density $\rho_{\rm atm}(r)$ as a fraction of $\rho(r)$ is adopted, 
the characteristic disk surface temperature $T_S(r)$ is determined iteratively by requiring 
$\int F_\nu(\rho_{\rm atm}(r),T_S(r)) d\nu = D(r)$ for a constant-density atmosphere, or
$\int E_\nu(\rho_{\rm atm}(r),T_S(r)) d\nu = D(r)$ for an exponential-density atmosphere.
Once $T_S(r)$ is known, the emergent spectrum as a function of $\nu$ is calculated as 
the area-weighted integral of $F_\nu$ or $E_\nu$ over the disk.
We present results for a range of atmospheric densities 
for both non-blackbody spectral models considered.

For comparison with microlensing and reverberation mapping observations,
we assume a face-on disk and calculate the flux-weighted and
response-function-weighted \citep{2007MNRAS.380..669C} mean radii, respectively:
\begin{equation}
r_f(\lambda)=\frac{\int_{r=r_{\rm in}}^{r=\infty} r ~X_\lambda(r) ~r~dr}{\int_{r=r_{\rm in}}^{r=\infty} X_\lambda(r) ~r~dr}
{\rm ~~and~~} 
r_r(\lambda)=\frac{\int_{r=r_{\rm in}}^{r=\infty} r ~\psi_\lambda(r) ~r~dr}{\int_{r=r_{\rm in}}^{r=\infty} \psi_\lambda(r) ~r~dr}
\end{equation}
where $X_\lambda(r)=B_\lambda(r)$ for a standard disk,
$F_\lambda(r)$ for a constant-density scattering atmosphere (\S \ref{conair}),
and $E_\lambda(r)$ for an exponential scattering atmosphere (\S \ref{expair}),
and $\psi_\lambda(r)$ is the response function
(for a derivation, see Horne et~al.\ in preparation).
We assume instantaneous thermalization and re-emission of incident illumination.
We refer to these wavelength-dependent radii collectively as $r(\lambda)$ and 
express them in light-days for comparison with time lag observations;  
the mean time lag is insensitive to disk inclination
(Fig.\ 1 of \citealt{2017ApJ...835...65S}).

\section{Comparison to NGC 5548 Observations} \label{ngc5548}

To compare with the observations of NGC 5548, 
we adopt a cosmology with $\Omega_M = 0.28$, $\Omega_\Lambda = 0.72$, and $H_0 = 70$ km s$^{-1}$ Mpc$^{-1}$ \citep{2011ApJS..192...18K}, 
which yields a luminosity distance of 74.5 Mpc at its redshift of $z = 0.017175$ \citep{2015ApJ...806..128D}.
We model a disk with viscosity parameter $\alpha=0.1$
around a non-rotating black hole of mass 
$M=(6.66 \pm 2.17)\times 10^7\,M_\odot$ \citep{agnstorm5Pei}. 
The Schwarzschild radius is 
$R_{\rm Sch}=1.97\times 10^{13}$~cm = 657 light-seconds
and the Eddington limit is $ L_{\rm Edd} = 4\pi GM\mu_ec/\sigma_T
= 9.79 \times 10^{45} {\rm \,erg~s}^{-1}$, where $\mu_e=2m_p/(1+X)$
is the mean mass per electron.
The corresponding Eddington mass accretion limit is 
$\dot{M}_{\rm Edd} = L_{\rm Edd}/(0.1\eta_{0.1}c^2) = 9.39 \eta_{0.1}^{-1} \times 10^{25}$ g\,s$^{-1}$; for our non-rotating black hole with $r_{in}=3R_{\rm Sch}$, we adopt the appropriate nonrelativistic value of $\eta_{0.1}=5/6$ \citep{fkr}.
NGC 5548 has bolometric luminosity $\simeq 10^{44.83}$ erg s$^{-1}$
\citep{WooUrry2002} and so we adopt $L/L_{\rm Edd}=0.069$.

We calculate $X_\lambda(r)$ for 
$3<r/R_{\rm Sch}<4000$ for a range of atmospheric densities
from $\log(\rho_{\rm atm}/\rho)=0$ to $-5$ for both scattering atmosphere models,
and then calculate $r(\lambda)$ for each.
The transition between radiation- and gas-pressure-dominated regions occurs at
$r \simeq 223 R_{\rm Sch}$,
as compared to $r \simeq 237 R_{\rm Sch}$ 
predicted by SS73 Eq.\ 2.17 or NT73 Eq.\ 5.9.9
for the NGC 5548 parameters adopted in this section.
The difference arises because the values of $T(r)$ and $\rho(r)$ 
in gas pressure-dominated regions differ slightly from what is assumed in the approximations 
used in deriving those equations (see the end of the Appendix).

In Figure \ref{f_time}, for comparison with Figure 5 of F16, we plot the
wavelength-dependent relative disk reverberation time lag, estimated from the
response-function-weighted mean radii $r(\lambda)-r$(1367~\AA) (solid lines)
for blackbody disks (black) and scattering-atmosphere disks with constant density (red) and exponential density (cyan), all with $L=0.069L_{\rm Edd}$.
At $\rho_{atm}/\rho=0$, the relative time lags lie above the blackbody disk value at
1000 \AA\ $\lesssim\lambda\lesssim$ 6000 \AA, and below it at other wavelengths.
As $\rho_{atm}/\rho$ decreases, the relative time lags decrease further at
$\lambda\lesssim$ 1000 \AA\ and increase at longer wavelengths until they are 
above the blackbody disk values at all wavelengths plotted (last three panels).
As expected from \S \ref{twomodels}, at sufficiently low $\rho_{\rm atm}$ the
time lags at each radius approach limiting values for an exponential-density atmosphere
and continue to increase for a constant-density atmosphere.

The dotted lines in Figure \ref{f_time} show results for flux-weighted mean radii, 
to illustrate the relative disk sizes expected in microlensing observations. 
Such radii are smaller than response-function-weighted radii
because the latter give high weight to radii
that have not reached the Rayleigh-Jeans limit at a given $\lambda$.

Scattering atmospheres with sufficiently low relative densities
can explain the observed wavelength-dependent reverberation signal in NGC 5548.
For response-function-weighted radii with $\log (\rho_{\rm atm}/\rho)$=$-4$,
for an exponential-density atmosphere 
we find $r_r(\lambda) \propto \lambda^{1.14\pm 0.02}$,
and for a constant-density atmosphere we find
$r_r(\lambda) \propto \lambda^{1.00\pm 0.02}$ at $1000<\lambda<4000$\,\AA\ and
$r_r(\lambda) \propto \lambda^{0.82\pm 0.02}$ at $\lambda>4000$\,\AA.

In Figure \ref{f_fnu}, we plot the flux density $f_\nu$ vs.\ $\nu$ from an annulus at 
60 $R_{\rm Sch}$,
to illustrate how the emergent spectrum changes with $\rho_{\rm atm}/\rho$ for both scattering atmosphere models.
The middle right panel includes color-corrected blackbody curves with peak wavelengths matching both models.
Because both models have more flux at short and long wavelengths than the matching color-corrected blackbody curves, modeling disk emission with the latter
(e.g., \citealt{2006ApJ...647..525D}; \citealt{2012MNRAS.420.1848D})
will not fully reproduce emission from these scattering atmosphere models.

In the left panel of Figure \ref{f_fnumJy}, 
we plot $f_\nu$ for NGC 5548 vs.\ expectations for disks with $\rho_{\rm atm}=10^{-4}\rho$.
The constant-density model overpredicts $f_\nu$ by less than a factor of three 
at the wavelengths plotted.  However, panel 5 of Figure \ref{f_sed} shows that
the integrated $\lambda F_\lambda$ for such a disk peaks at much
shorter wavelengths than for blackbody disks and is a poor match to the 
optical$-$UV$-$X-ray SED of NGC 5548 \citep{2015A&A...575A..22M}.
Note that panel (1) of Figure \ref{f_sed} shows the expected deviation 
of even a standard thin disk's SED from a sum of blackbodies,
as seen in Figure 5.10.1 of NT73 and Figure 3 of SS73.
However, this deviation is due almost entirely to emission from radii $r<20 R_{\rm Sch}$.
The differences are small at larger radii;
e.g., the SED from $r>60 R_{\rm Sch}$ is shown in panel (1).

For objects where such SEDs conflict with observations,
winds from the disk may reduce $\dot{M}$ at small radii and thus
the luminosity from the inner disk, preventing the SED from peaking at
too short a wavelength (\citealt{2012MNRAS.426..656S,2014MNRAS.438.3024L};
see also \citealt{proga05}).
Alternatively, at small radii the disk may transition to a
thick torus whose emission peaks in the far or extreme ultraviolet 
\citep[e.g., \S 2 of][]{gd16}.
As a purely illustrative limiting case of the above,
in the right panel of Figure \ref{f_fnumJy}
and the lower left panel of Figure \ref{f_sed} we show SEDs for models with
$\log(\rho_{\rm atm}/\rho)=-4$ but from disk radii $r>60R_{\rm Sch}$ only ($r>11$ light-hours).
These SEDs show the relative fluxes of the inner and outer disks at those wavelengths.
As shown in Figure \ref{f_timedashed} (left panel), our model's predicted mean 
relative disk reverberation time lags at $\lambda>1367$\,\AA\ (solid lines) are
not greatly changed if the emission from $r<60R_{\rm Sch}$ is excluded (dashed lines).
Absolute time lag predictions, in contrast,
are strongly dependent on the UV emission properties of the inner disk.
Figure \ref{f_timedashed} (right panel) shows that our model's predicted mean {\em absolute}
disk reverberation time lags with emission from $r<60R_{\rm Sch}$ excluded (dashed lines)
are closer to the AGN STORM observations of NGC 5548 than our model's predictions including
the full disk (solid lines).  
Thus, models that suppress optical/UV emission disk from radii $r<60R_{\rm Sch}$ 
to better fit the constraints of observed AGN SEDs 
may be able to simultaneously match inferred relative and absolute disk sizes.

\section{Discussion} \label{theend}

If the atmospheric density in an AGN accretion disk is sufficiently low, 
scattering in the atmosphere can produce a non-blackbody emergent spectrum.
Disks with these scattering atmospheres
have lower brightness temperatures
and higher color temperatures at ultraviolet and optical wavelengths
than the blackbody temperature of a disk of the same bolometric luminosity;
thus, they can in principle reconcile disk theory sizes and flux sizes
with microlensing sizes \citep[see, e.g., \S1 of][]{ad}.

The AGN STORM observations of NGC 5548 are 
in many --- but not all --- aspects consistent with a disk with a constant-density 
scattering atmosphere with density $\rho_{\rm atm}=10^{-4}\rho$.
In particular, although the observed optical$-$UV$-$X-ray SED of NGC 5548 is not well matched
by that model (Figure \ref{f_sed}), the observed optical$-$UV flux from NGC 5548 is lower than
in all models at our adopted Eddington ratio of $L=0.069L_{\rm Edd}$ (Figure \ref{f_fnumJy}).
That Eddington ratio is based on the bolometric luminosity from \cite{WooUrry2002}, which was
found by summing multiwavelength flux values for NGC 5548 from the literature.
Contamination of flux measurements by the host galaxy, particularly at infrared wavelengths,
could mean that the true Eddington ratio is lower than we have assumed, 
which would help explain the lower optical$-$UV flux level (e.g., \cite{gd16}
adopt $L=0.040L_{\rm Edd}$).

As a consistency check, we note that the physical density corresponding to
$\rho_{\rm atm}=10^{-4}\rho$ in our NGC 5548 disk models
is $10^{-10}$-$10^{-14}$ g cm$^{-3}$ at 4-4000 $R_{\rm Sch}$ from the black hole,
with a minimum at 8.5 $R_{\rm Sch}$.
The vertical length scale corresponding to $\tau_{\rm es}=1$ at such densities is
$h_{\rm es}$=$10^{10}$-$10^{14}$ cm.
The atmospheric scale height/radius ratio $h_{\rm es}/r$ peaks at $0.18$ at 8.5 $R_{\rm Sch}$
and has a value $h_{\rm es}/r < 10^{-2}$ at $40 < r/R_{\rm Sch} < 4000$.
A self-consistent atmosphere model must have $h_{\rm es}/r \ll 1$,
so our constant-density model results at $r<40 R_{\rm Sch}$ should be treated with caution.
For an exponential atmosphere with $\rho_{\rm atm}=10^{-4}\rho$ and scale height $h_1$
(Eq.\ \ref{eq_h1}),
we find at all radii that $h_1/r \lesssim 6 \times 10^{-5}$, so $h_1 < h_{\rm es}$.
Thus, the semi-infinite exponential atmosphere model we use (\S \ref{expair}) 
is a reasonable approximation.  Compared to an infinite exponential atmosphere,
it neglects a region of optical depth only $\tau = \kes\rho_{\rm atm} h_1 < 1$. 

We can compare our $\log (\rho_{\rm atm}/\rho)$=$-4$ 
model predictions to the results of microlensing observations.
For flux-weighted radii with $\log (\rho_{\rm atm}/\rho)$=$-4$,
for an exponential-density atmosphere
we find $r_f(\lambda) \propto \lambda^{1.09\pm 0.02}$,
and for a constant-density atmosphere we find
$r_r(\lambda) \propto \lambda^{0.84\pm 0.02}$.
In the Einstein Cross, \cite{2008A&A...490..933E} found flux-weighted radii
$r_f \propto \lambda^{1.2\pm 0.3}$.
In the quasar HE 1104$-$1805 \cite{2015ApJ...798...95B} found 
$r_f \propto \lambda^{1.0(+0.3,-0.56)}$ and a size larger than expected by a factor of 4 at
1900\,\AA, as compared to our prediction of a factor $\sim 3$ (Figure \ref{f_relativesize}).
In a sample of 11 quasars, \cite{2011ApJ...729...34B} found 
$r_f \propto \lambda^{0.17\pm 0.15}$ and a size larger than expected by a factor of 
$10^{+7}_{-5}$ at 1736\,\AA, as compared to our prediction of a factor $\sim 3$.
The results from first two studies above are reasonably consistent with both the 
theoretical prediction of $r_f \propto \lambda^{4/3}$ and with our model results; 
see \cite{ad} for further discussion of the \cite{2015ApJ...798...95B} results.

For $\rho_{\rm atm}$ values low enough for scattering to help explain the inferred large sizes of AGN accretion disks, some mechanism must support the low-density atmosphere above the disk; magnetic pressure support is an obvious possibility.
\cite{2006ApJ...645.1402B} have shown that accounting for magnetic pressure 
yields a more extended accretion disk atmosphere that shifts the emergent SED
in a similar manner to our Figure \ref{f_sed}; see also
\cite{2007MNRAS.375.1070B}, 
\cite{2009ApJ...703..569D}, 
and \cite{2013ApJ...770...55T}. 
Such studies found photospheric densities 
of $\rho_{\rm atm}\simeq 10^{-3}\rho$
(Figure 6 of \citealt{2006ApJ...645.1402B})
and $\rho_{\rm atm}\simeq 2 \times 10^{-3}\rho$
(Figure 3 of \citealt{2009ApJ...703..569D})
in disks around stellar-mass black holes.
It remains to be seen if disks around supermassive black holes can reach the
lower atmospheric densities required for scattering atmospheres
to explain large apparent AGN disk sizes.

The models discussed here predict that scattering atmospheres of different densities will 
yield different apparent disk sizes and $t_{\rm delay}(\lambda)\propto\lambda^x$ slopes $x$.
In addition, at low atmospheric densities, these models predict
sizes relative to blackbody disks that have maxima at 
$\lambda$$\lesssim$1000\,\AA\ and decrease toward approximately unity
at longer wavelengths (Figure \ref{f_relativesize}).
Overall, disk sizes are predicted to be more anomalous
in the ultraviolet than in the optical,
and response-function-weighted radii to be more discrepant
from their corresponding blackbody values than flux-weighted radii, 
by up to a factor of two and a half 
even at the density of a standard thin disk (Figure \ref{f_real}).
However, we have not accounted for specific features in the bound-free opacity
such as the temperature-dependent iron opacity bump
(see \citealt{2016ApJ...827...10J} and references therein)
or the Balmer and Paschen edges.  In particular,
if the disk atmospheres contain sufficient \ion{H}{1}, disk sizes and time lags
at the wavelengths of \ion{H}{1} edges may differ noticeably from our predictions.  It remains
to be seen whether or not such an effect can explain the discrepant $U$-band lags noted in F16.

If scattering atmospheres are important in AGN disks, dependences of the
scattering atmosphere properties on AGN parameters such as $M$ and $\dot{M}$
may help explain why SDSS quasars do not show the expected trends of continuum
color or emission line properties with such parameters
\citep{2007ApJ...659..211B,2013ApJ...770...30B}. 

\begin{figure}
\includegraphics[scale=0.50]{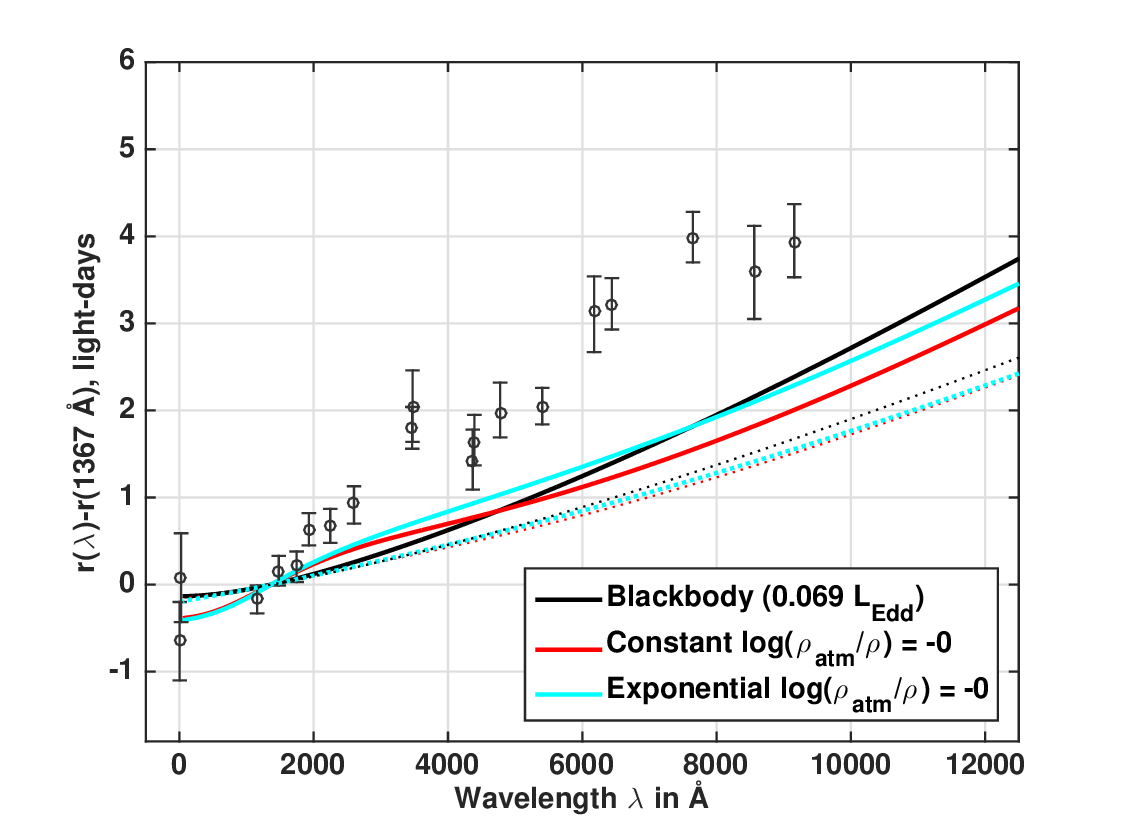}\includegraphics[scale=0.50]{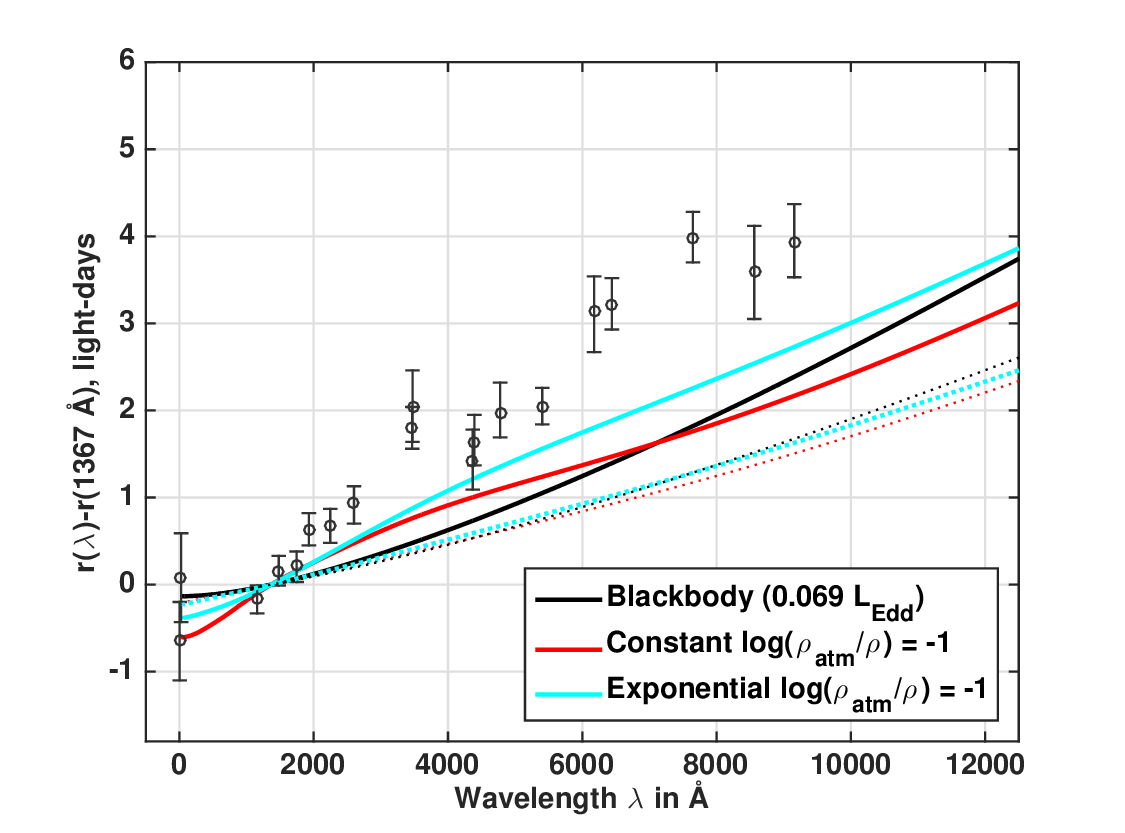}
\includegraphics[scale=0.50]{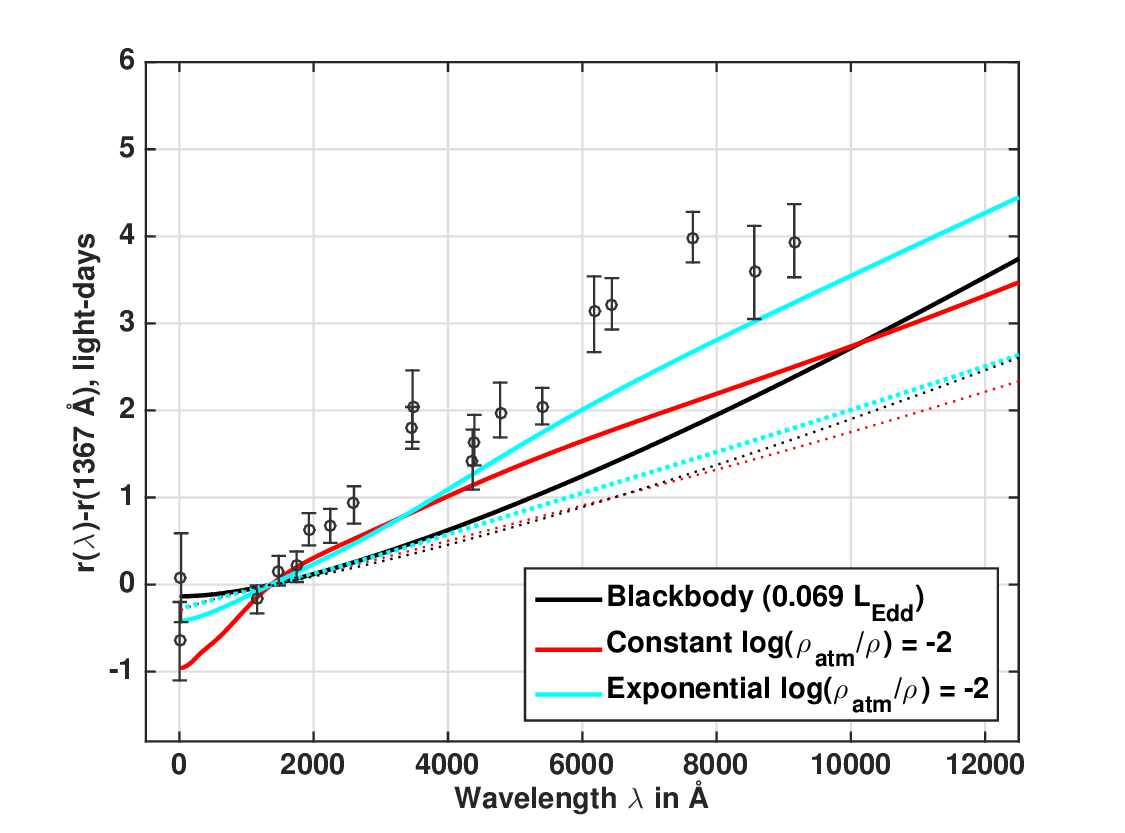}\includegraphics[scale=0.50]{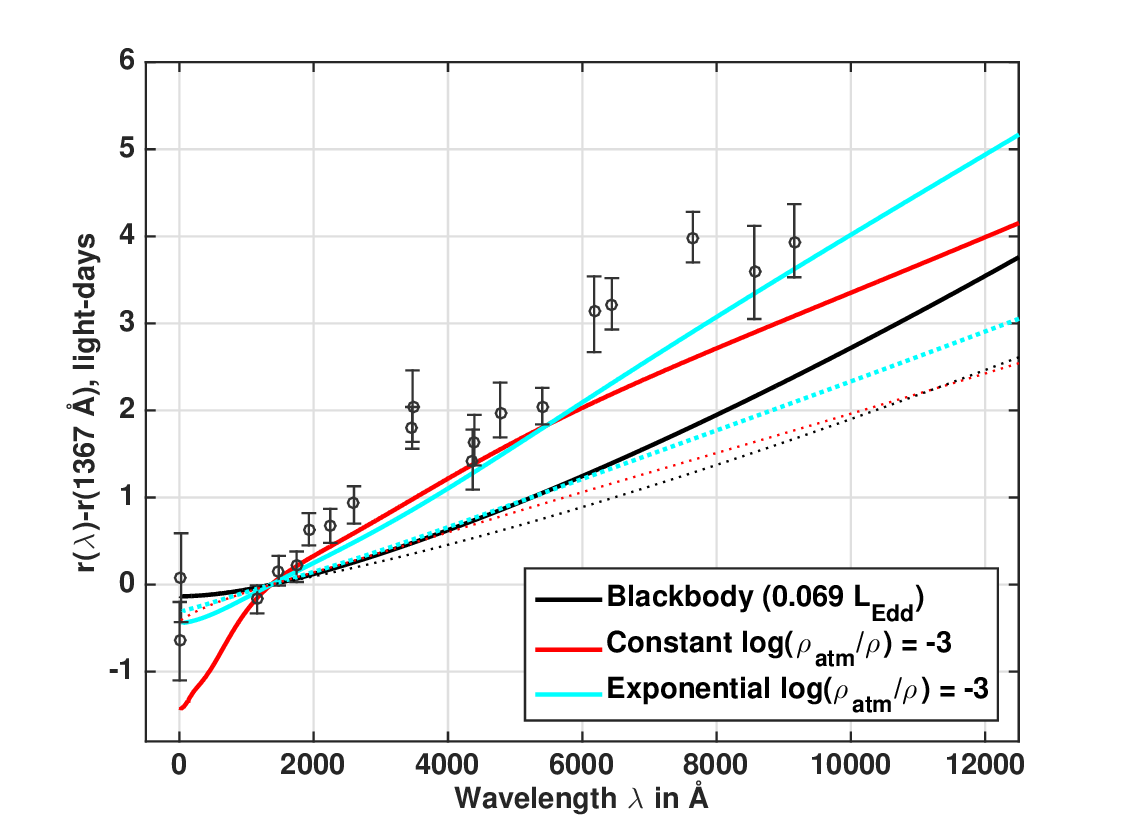}
\includegraphics[scale=0.50]{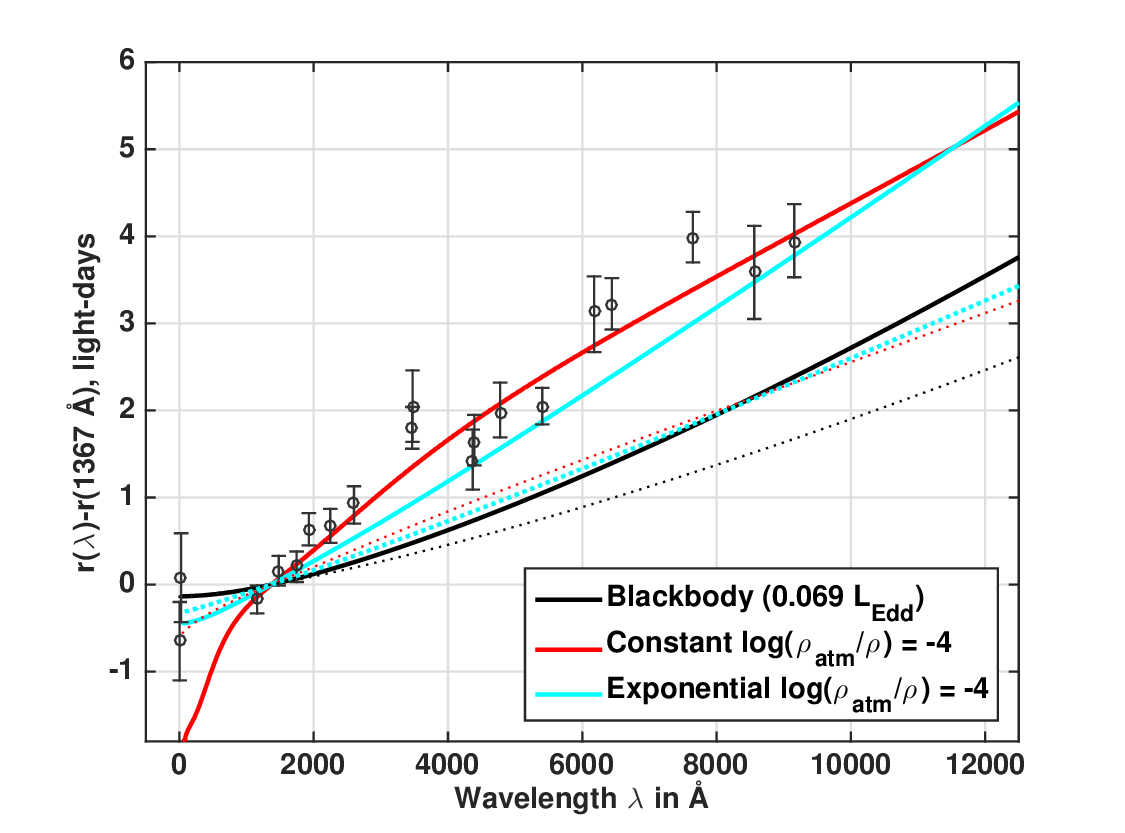}\includegraphics[scale=0.50]{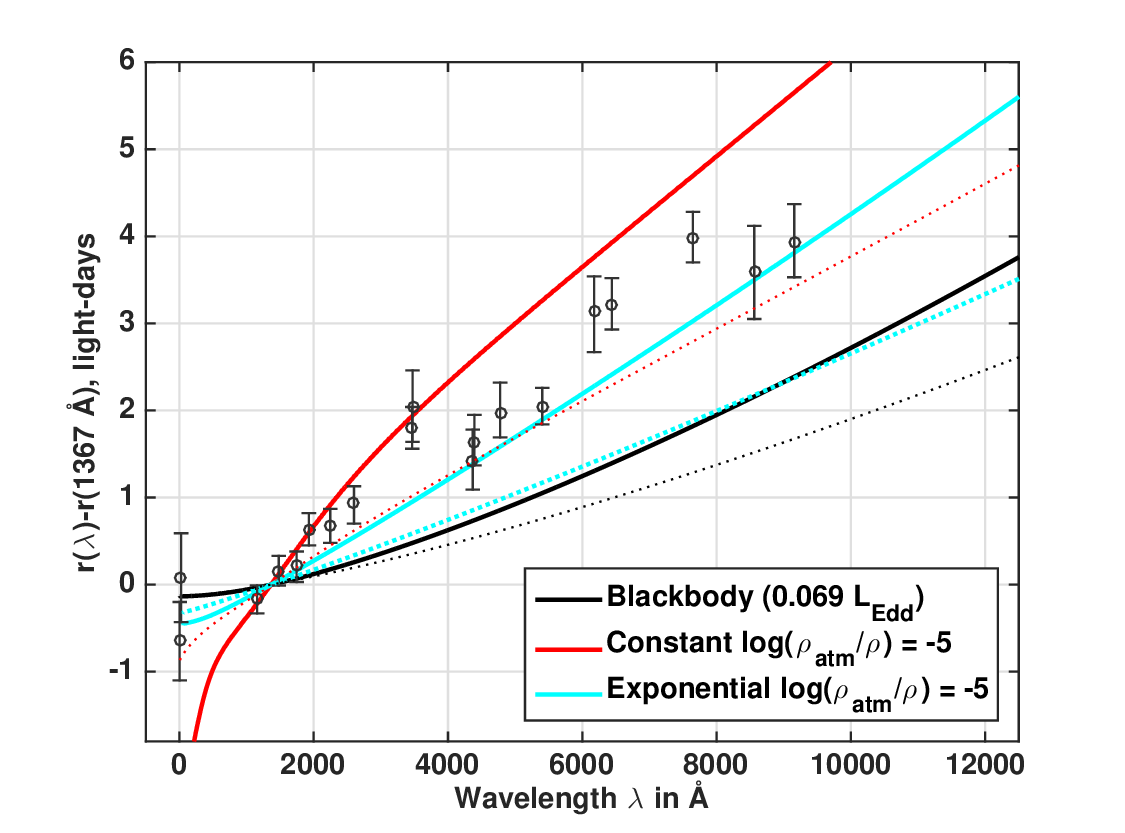}
\caption{The wavelength-dependent relative disk reverberation time lag estimated from the
response-function-weighted mean radius $r(\lambda)-r$(1367~\AA) (solid lines), 
overplotted on the wavelength-dependent time lag data for NGC 5548
(points and error bars, from F16; 
note that the $U$-band points just shortward of 4000~\AA\ are known to be discrepant).
Dotted lines show results for flux-weighted mean radii to illustrate the
relative disk sizes expected in microlensing observations. 
We plot curves for blackbody disks (black) and scattering-atmosphere disks with 
constant density (red) and exponential density (cyan), all with $L=0.069L_{\rm Edd}$.
The density of the scattering atmosphere relative to the average interior density is 
$\log (\rho_{atm}/\rho)=0$ and $-1$ (top), $-2$ and $-3$ (middle), and $-4$ and $-5$ (bottom).
At $\rho_{atm}/\rho=0$, the relative time lags lie above the blackbody disk value at 
1000 \AA\ $\lesssim\lambda\lesssim$ 6000 \AA, and below it at other wavelengths.
As $\rho_{atm}/\rho$ decreases, the relative time lags decrease further at
$\lambda\lesssim$ 1000 \AA\ and increase at longer wavelengths until they are
above the blackbody disk values at all wavelengths plotted (last three panels).
\label{f_time}}
\end{figure}

\begin{figure}
\includegraphics[scale=0.50]{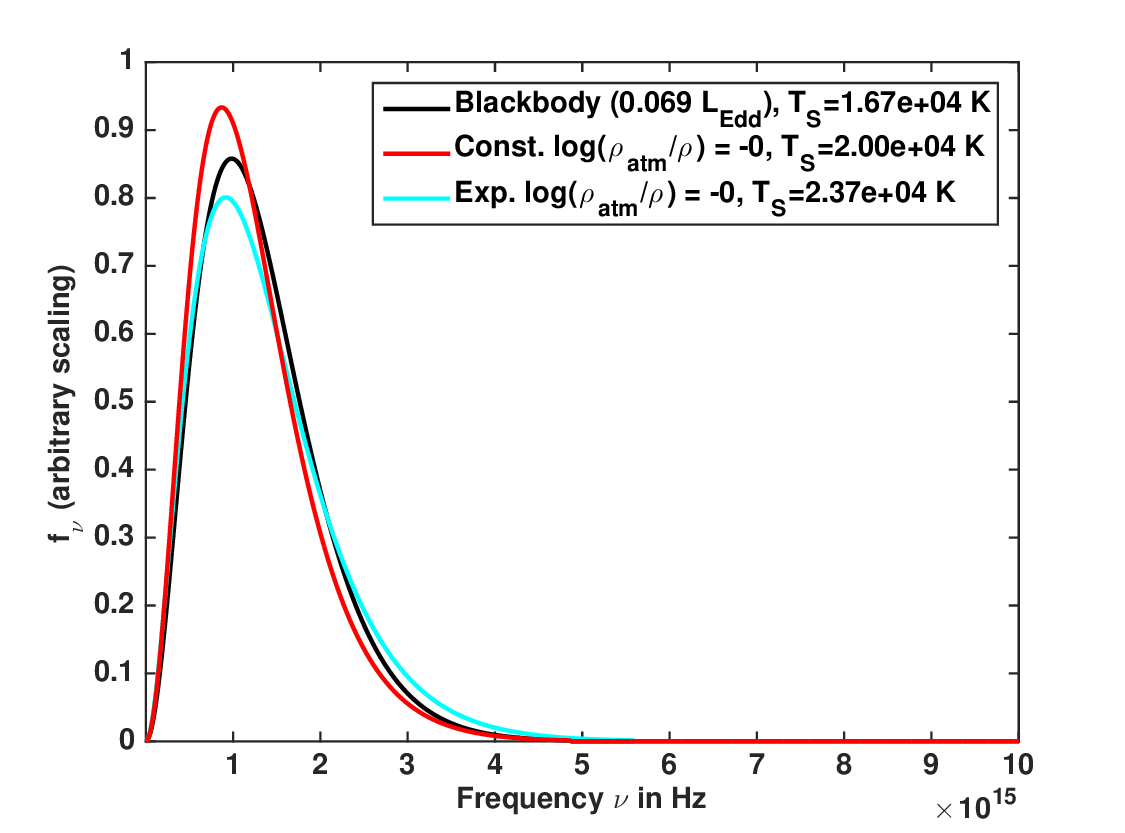}\includegraphics[scale=0.50]{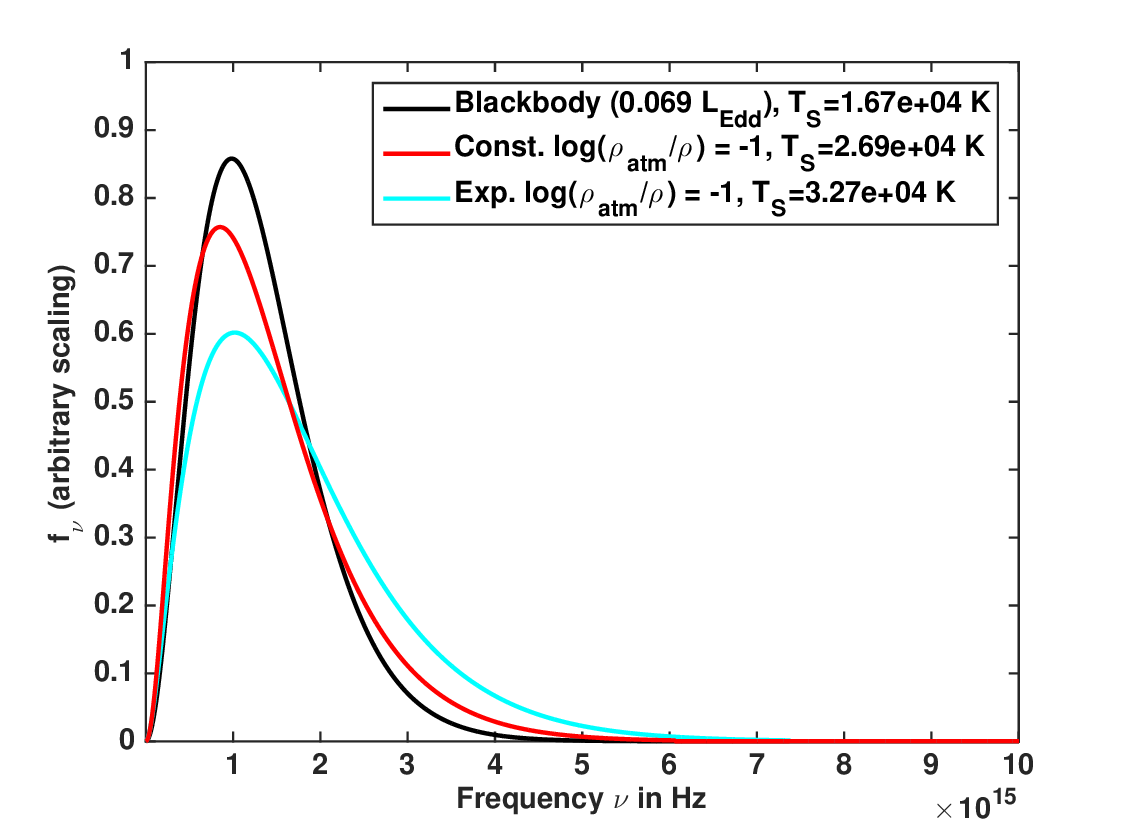}
\includegraphics[scale=0.50]{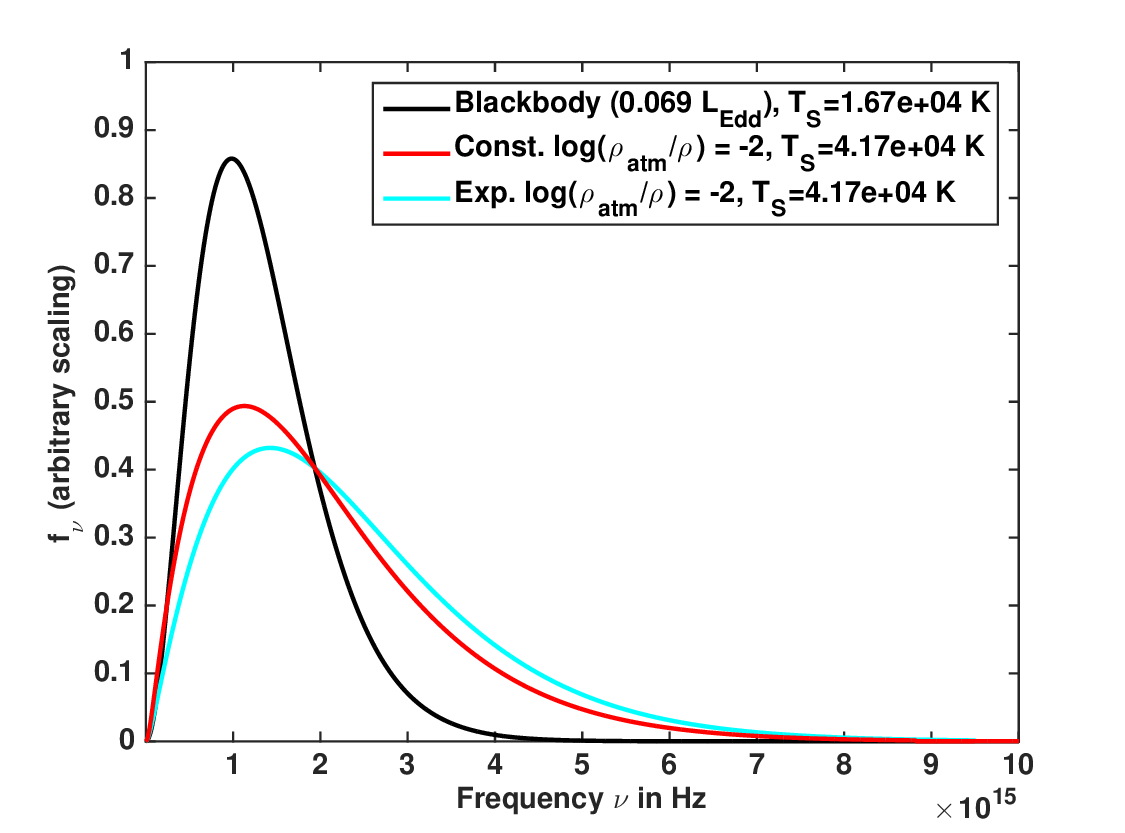}\includegraphics[scale=0.50]{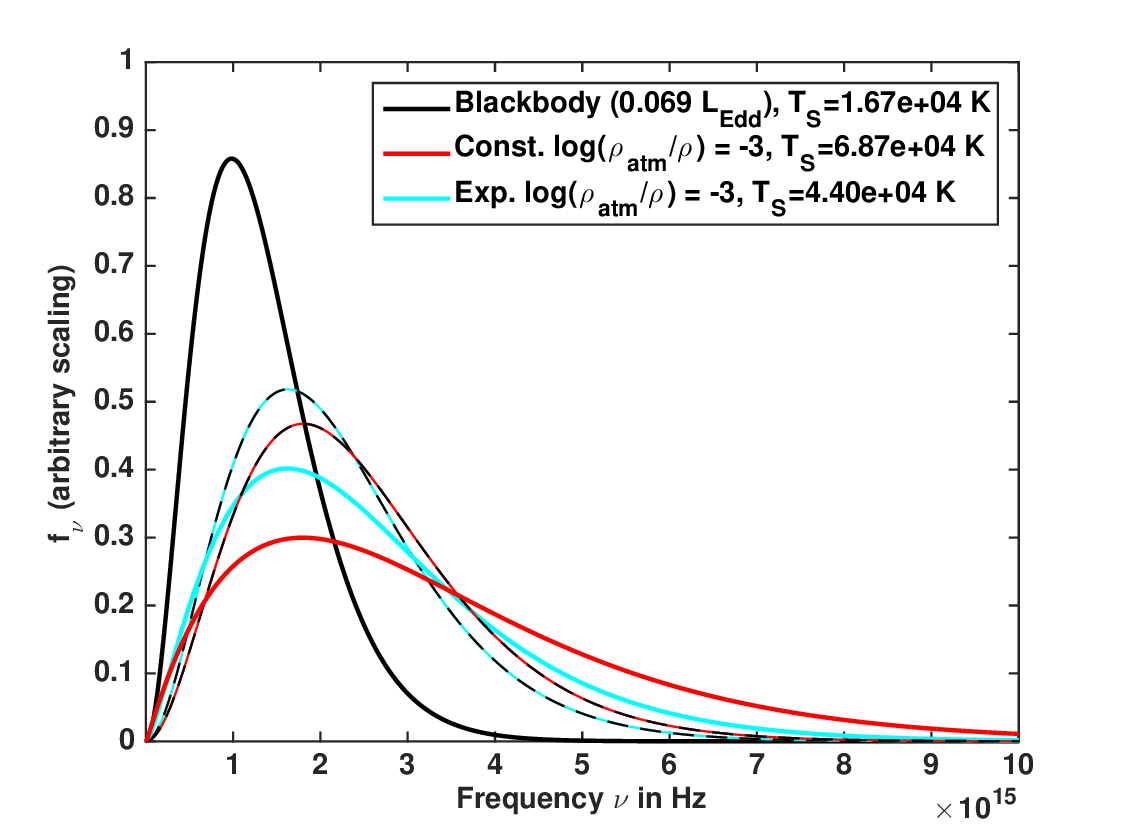}
\includegraphics[scale=0.50]{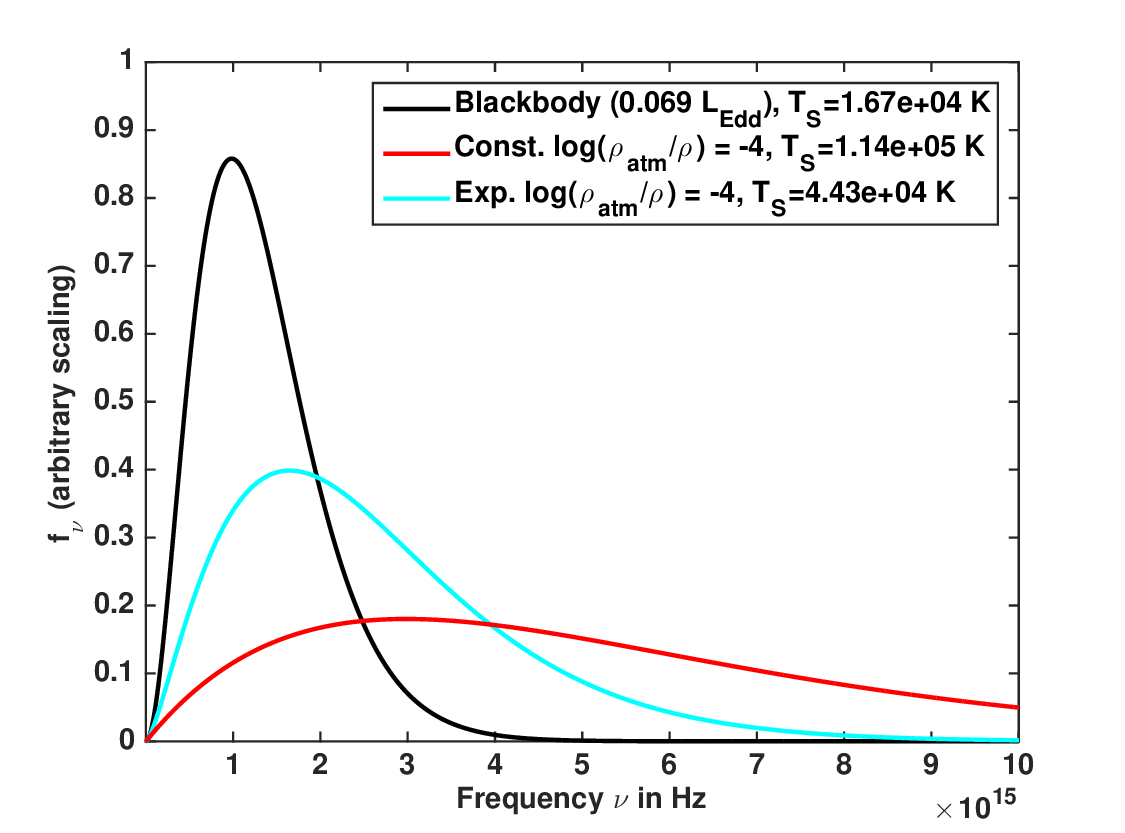}\includegraphics[scale=0.50]{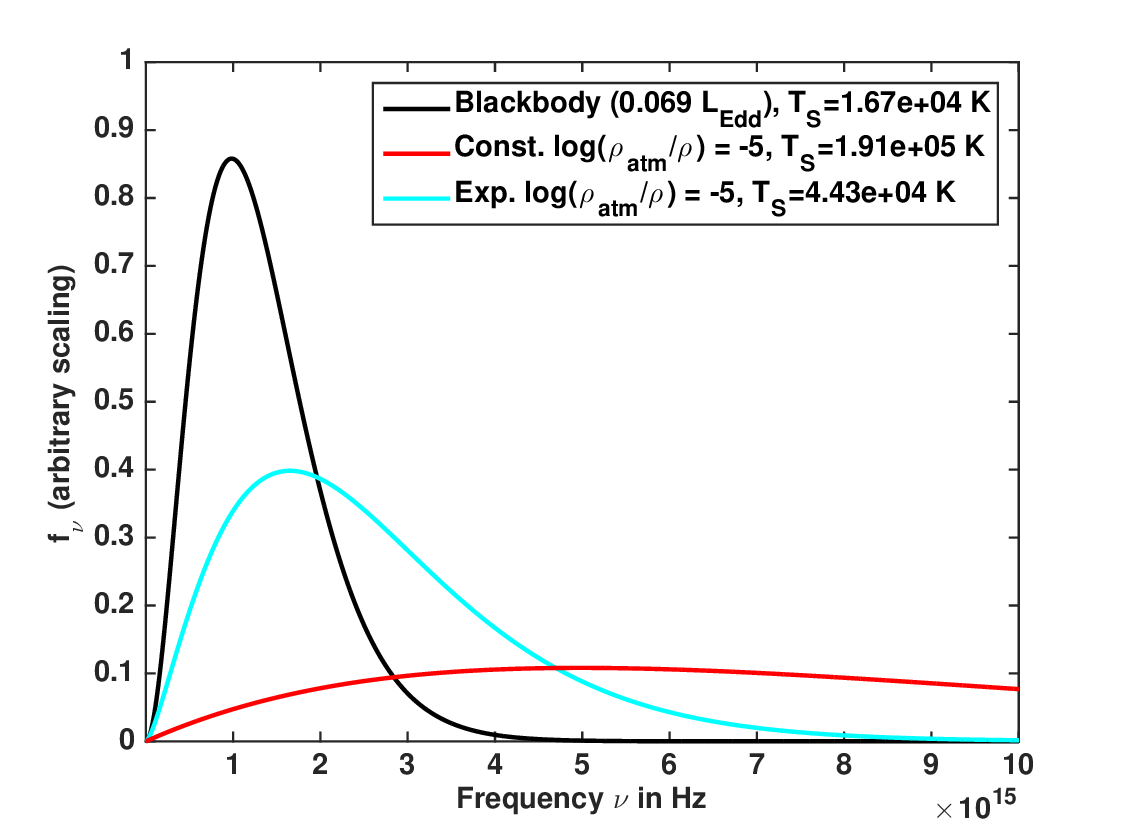}
\caption{Flux density $f_\nu$ from an annulus at 60 $R_{\rm Sch}$
in a disk with $L=0.069L_{\rm Edd}$ around NGC 5548.
Panel layout and color key are as in Figure \ref{f_time}.
The middle right panel includes color-corrected blackbody curves with the same peak wavelengths as the constant-density and exponential models 
(dashed red-black with $T_S=3.07$e+04 K and dashed cyan-black with $T_S=2.77$e+04 K, respectively).
\label{f_fnu}}
\end{figure}

\begin{figure}
\includegraphics[scale=0.50]{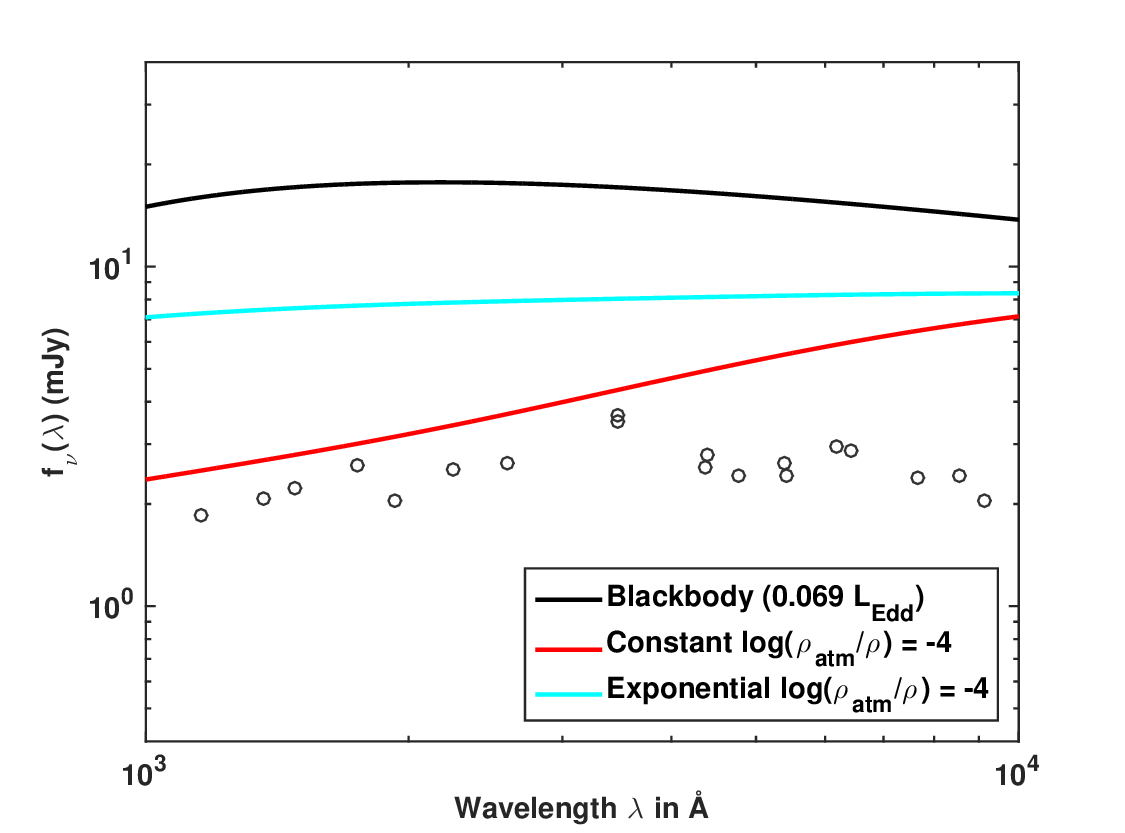}\includegraphics[scale=0.50]{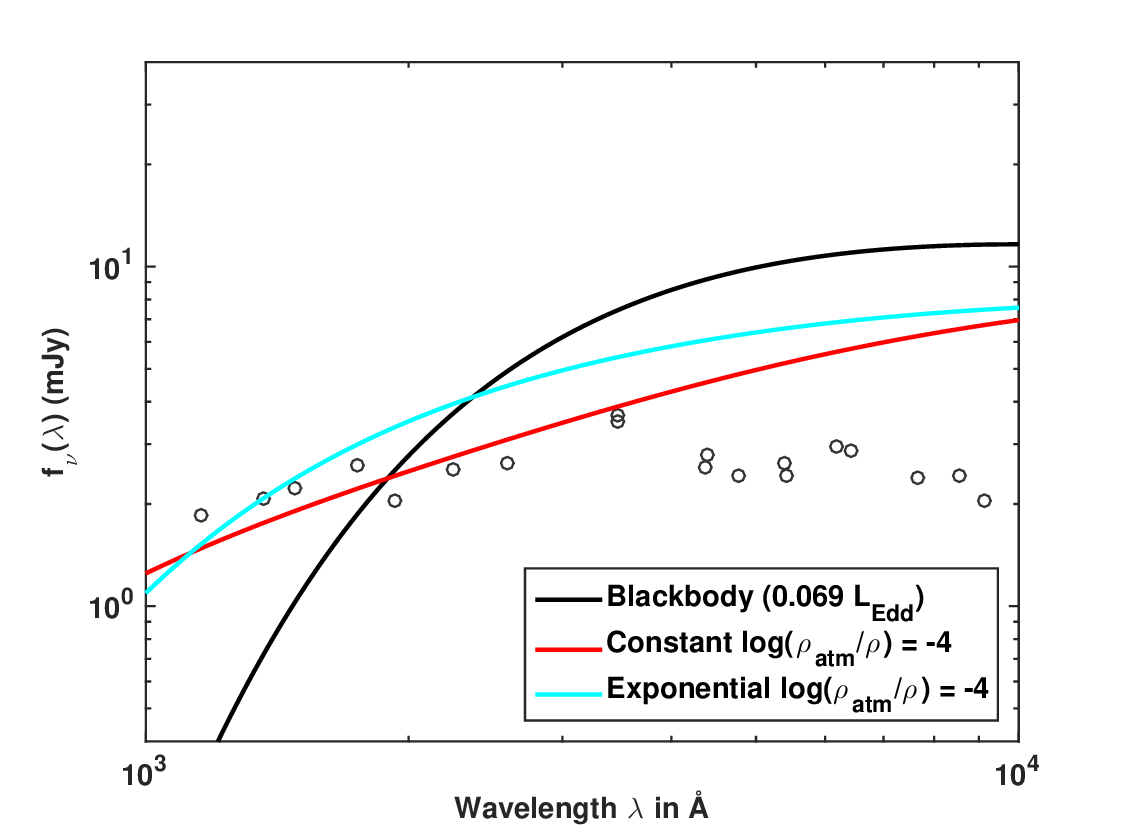}
\caption{In comparison to the average observed
$f_\nu$ values from Figure 11 of \cite{2017ApJ...835...65S},
we plot the predicted flux density from face-on accretion disks with
$L=0.069L_{\rm Edd}$ and $\log (\rho_{\rm atm}/\rho) = -4$
in NGC 5548, with the same color key as Figure \ref{f_time}.
The left panel shows $f_\nu$ from the entire disk; the right shows $f_\nu$ from
$r>60 R_{\rm Sch}$ ($r>11$ light-hours) only.
\label{f_fnumJy}}
\end{figure}

\begin{figure}
\includegraphics[scale=0.50]{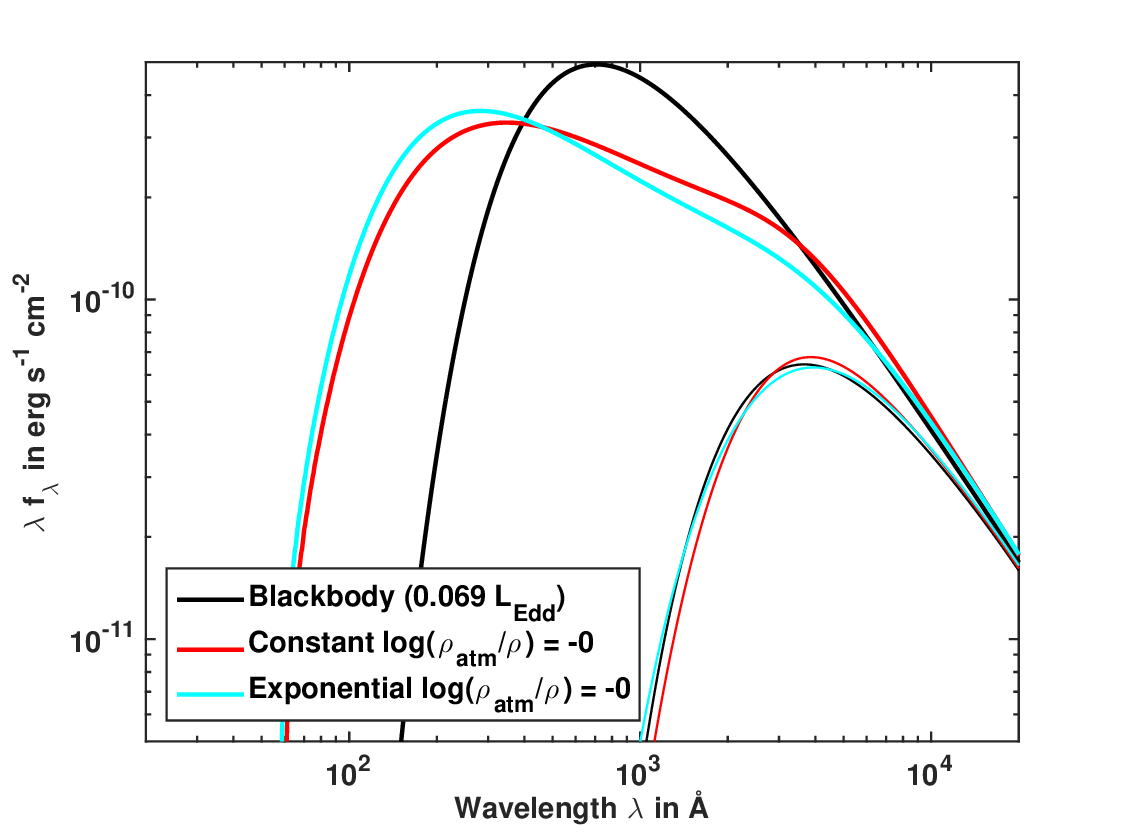}\includegraphics[scale=0.50]{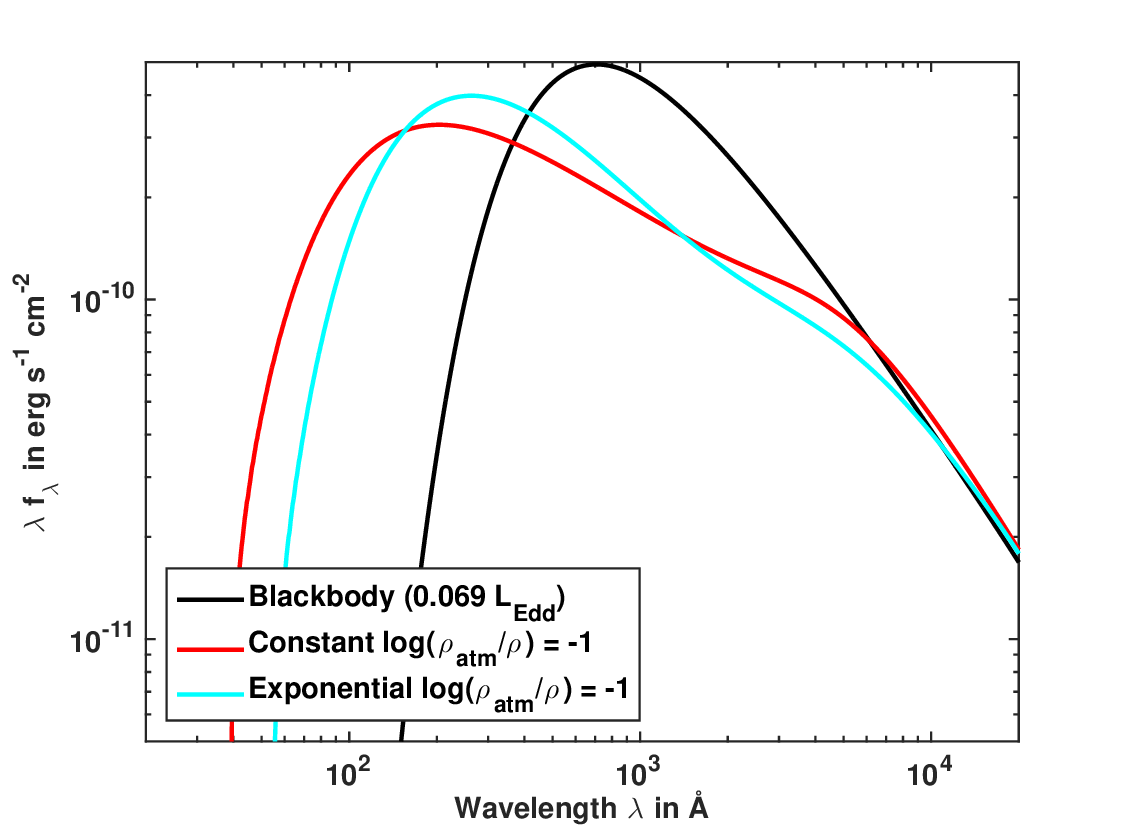}
\includegraphics[scale=0.50]{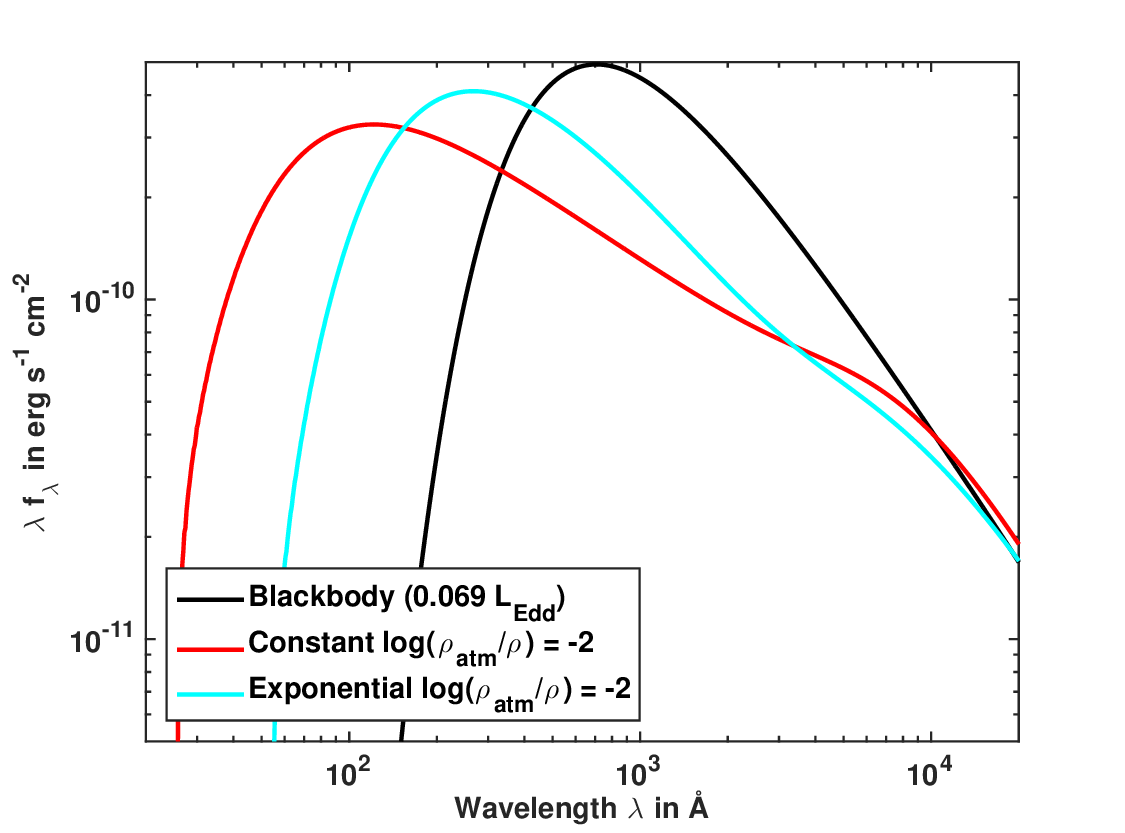}\includegraphics[scale=0.50]{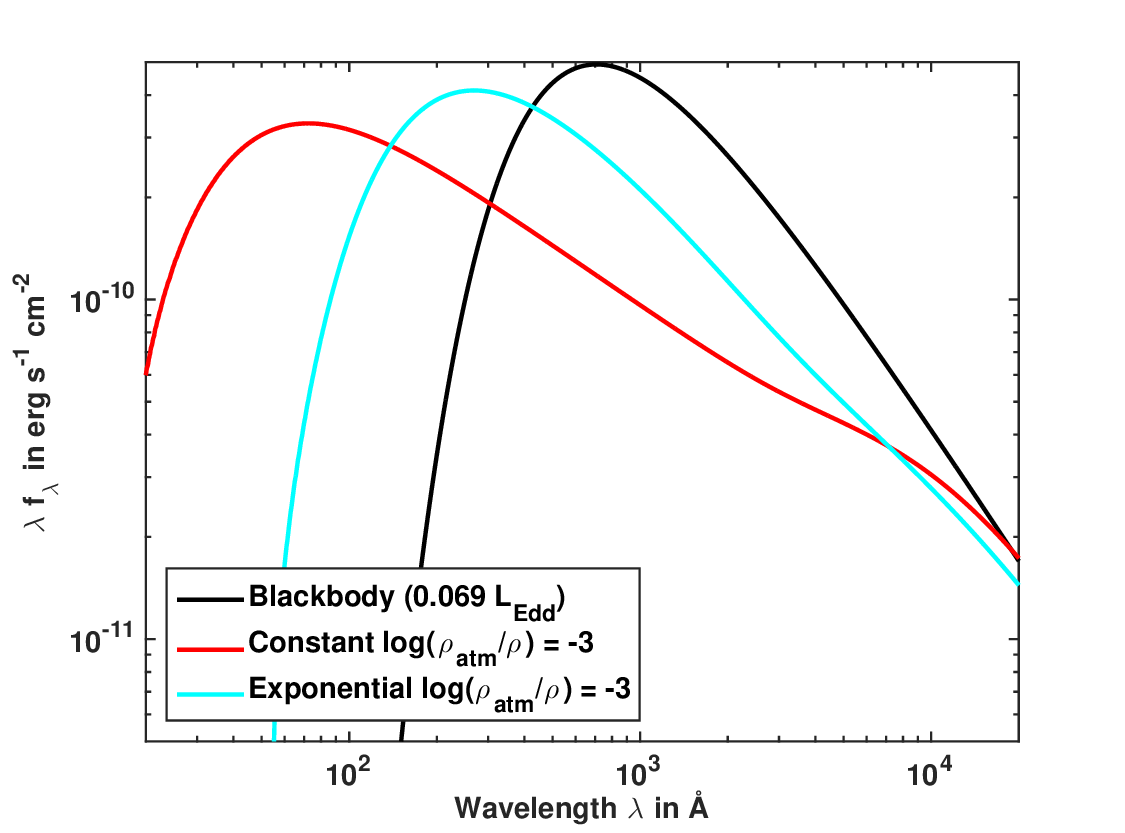}
\includegraphics[scale=0.50]{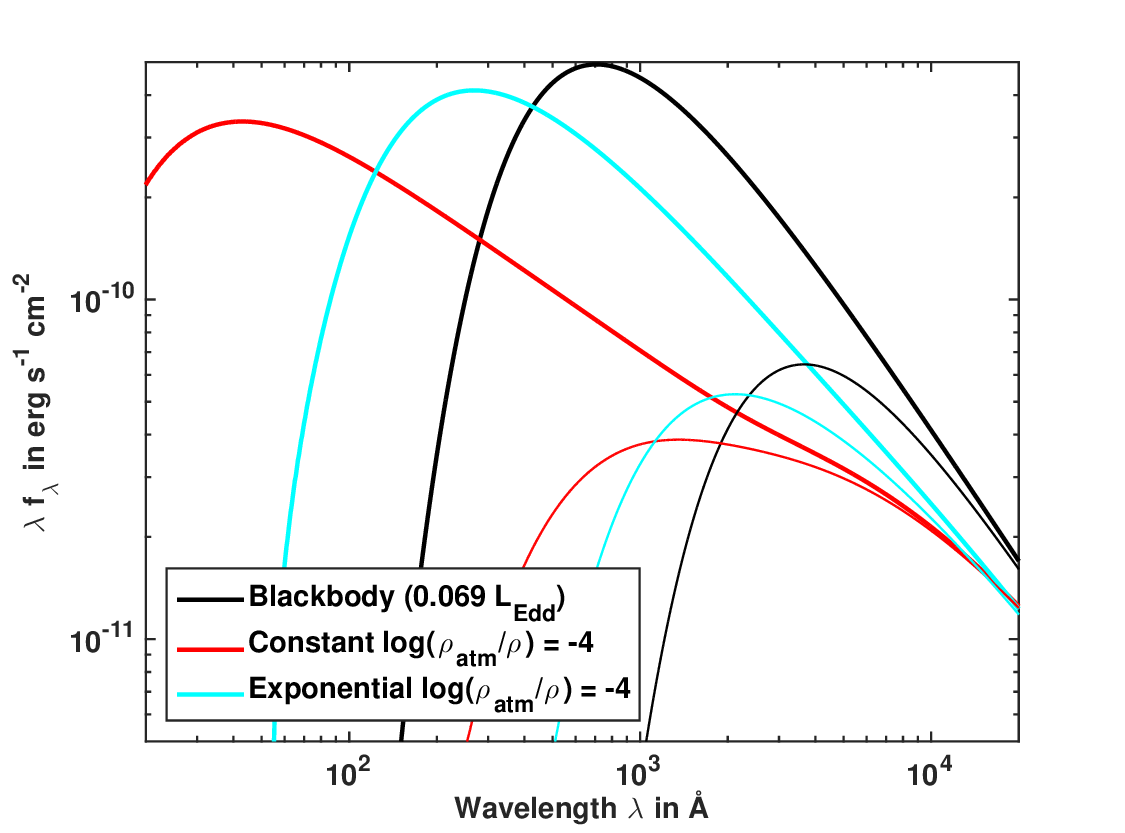}\includegraphics[scale=0.50]{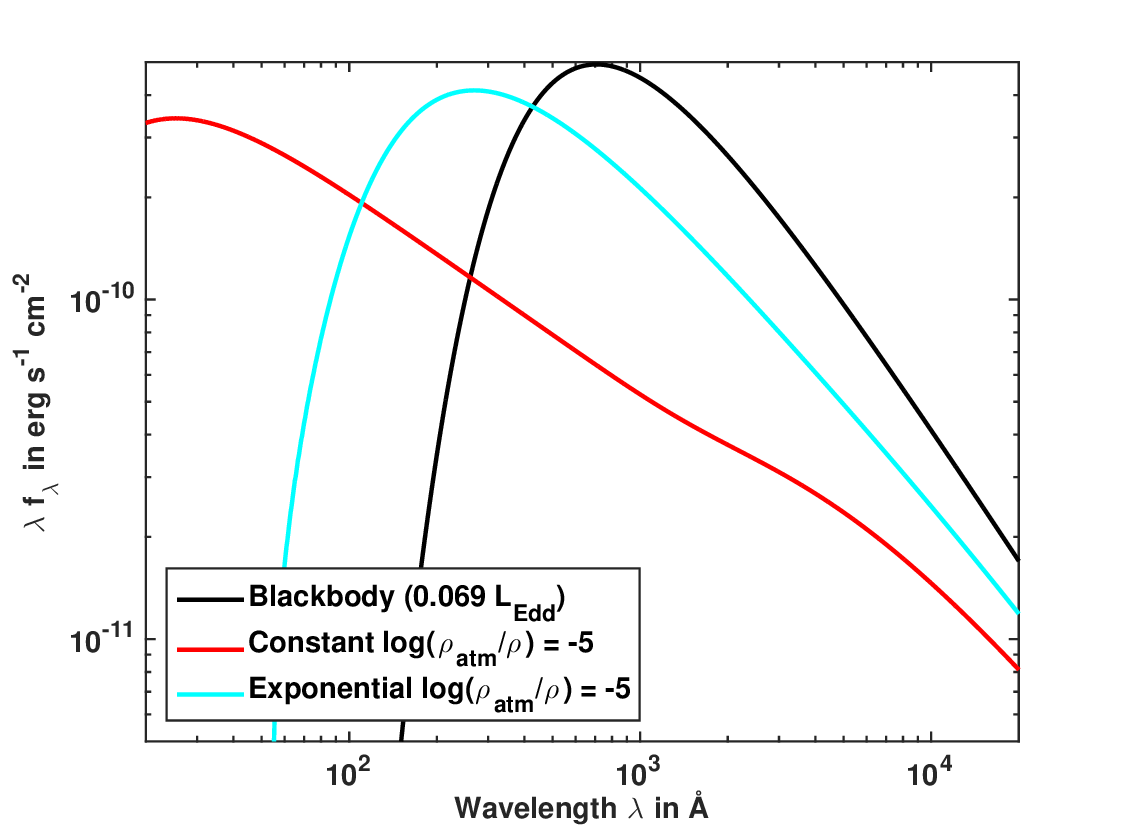}
\caption{The predicted observed spectral energy distributions (SEDs)
from face-on accretion disks with $L=0.069L_{\rm Edd}$ in NGC 5548.
Panel layout and color key are as in Figure \ref{f_time}.
The thin curves in the 
upper left and lower left panels
show $f_\nu$ from $r>60 R_{\rm Sch}$ only.
\label{f_sed}}
\end{figure}

\begin{figure}
\figurenum{5} 
\includegraphics[scale=0.50]{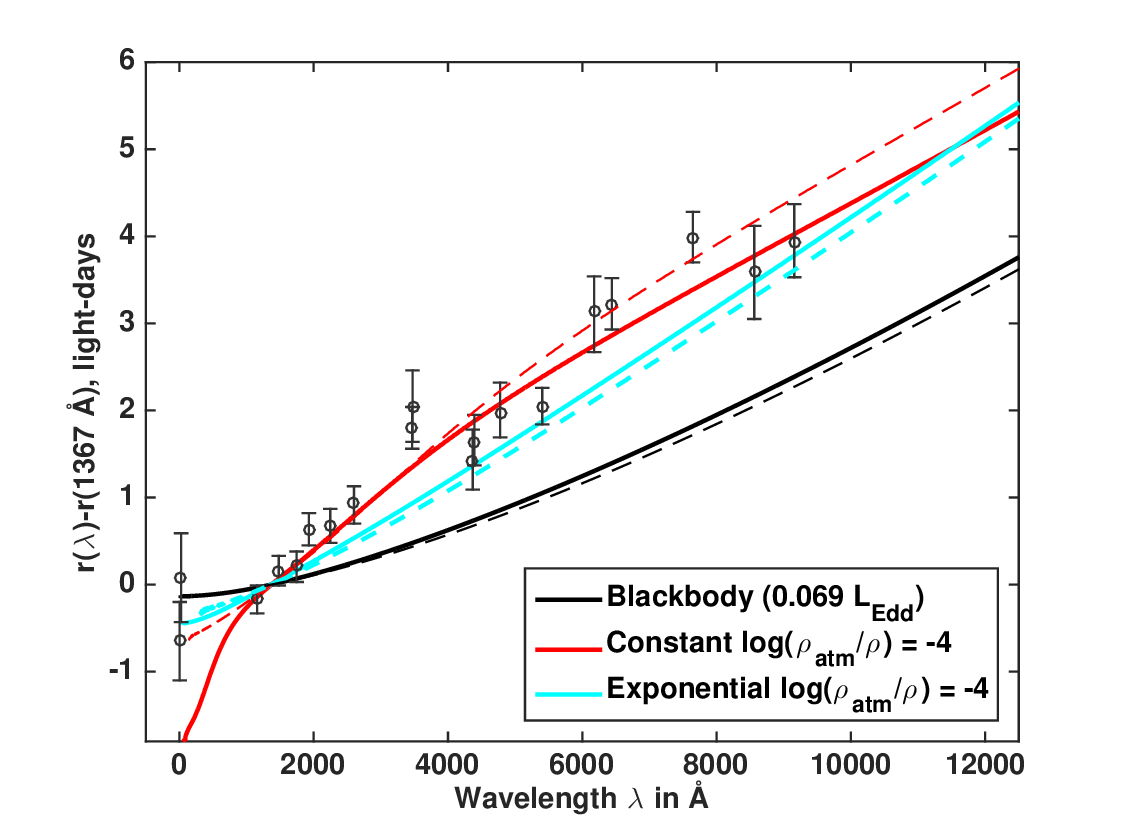}
\includegraphics[scale=0.50]{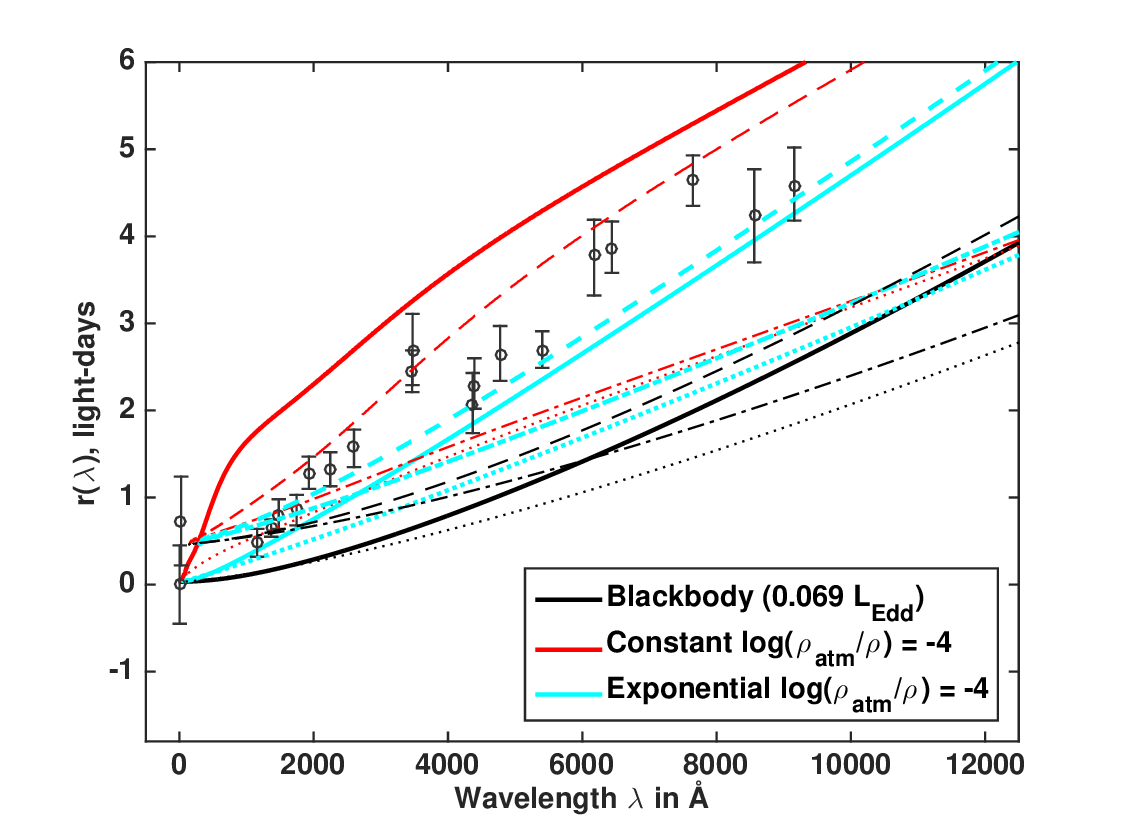}
\caption{Left panel: 
in the same fashion as Figure 1 (see that caption for details), we plot
the wavelength-dependent relative disk reverberation time lag estimated from the
response-function-weighted mean radius $r(\lambda)-r$(1367~\AA) (solid lines)
for blackbody disks (black) and scattering-atmosphere disks with
constant density (red) and exponential density (cyan), all with $\log (\rho_{atm}/\rho)=-4$.
Dashed lines show the results considering disk radii $r>60R_{\rm Sch}$ only.
Right panel: 
response-function-weighted mean radii $r(\lambda)$ (absolute disk reverberation 
time lags) for blackbody disks (black) and scattering-atmosphere disks with
constant density (red) and exponential density (cyan), all with $\log (\rho_{atm}/\rho)=-4$.
Dashed lines show the results considering disk radii $r>60R_{\rm Sch}$ only.
Dotted and dotted-dashed lines show flux-weighted radii for the full disk
and for disk radii $r>60R_{\rm Sch}$ only, respectively.
Points from F16 are plotted assuming zero time delay for the {\em Swift} hard X-ray band.
\label{f_timedashed}}
\end{figure}

\begin{figure}
\figurenum{6} 
\includegraphics[scale=0.65]{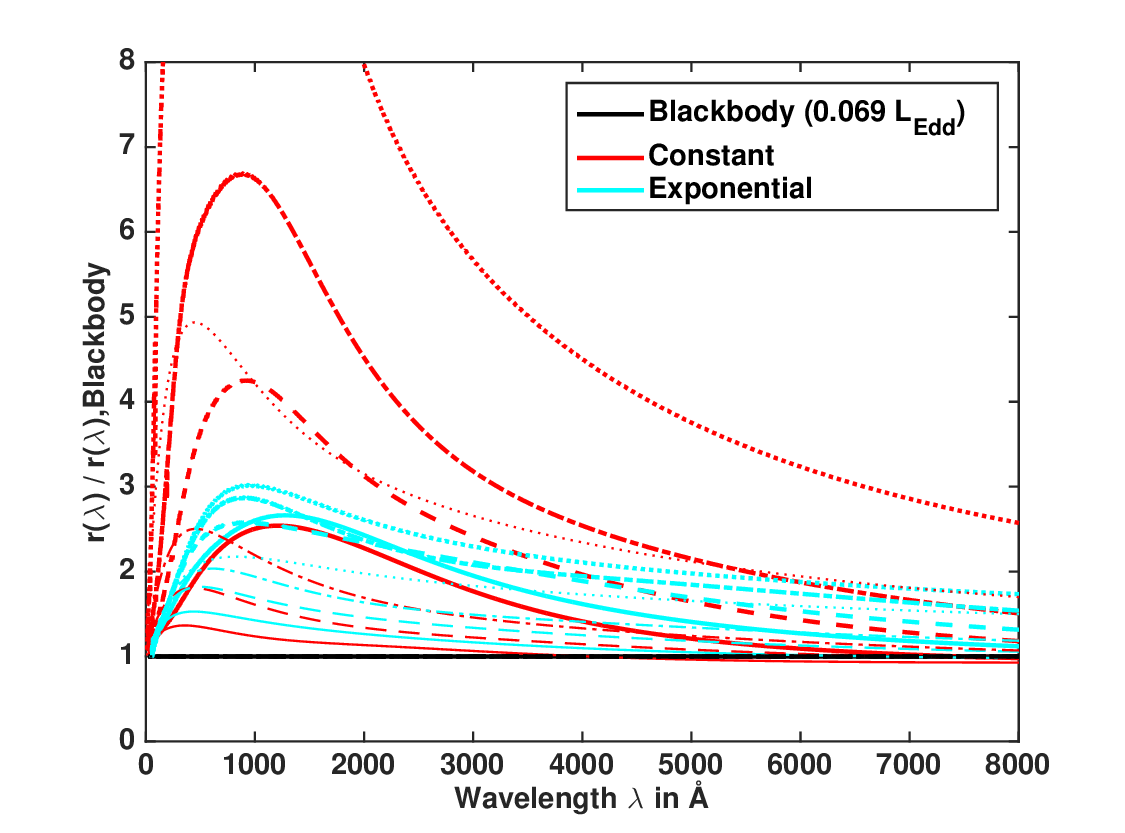}
\caption{Mean radii $r(\lambda)$ divided by the corresponding $r(\lambda)$ for 
a blackbody disk, for scattering-atmosphere disks with constant density (red)
and exponential density (cyan), all with $L=0.069L_{\rm Edd}$.
The curves shown are for $\log (\rho_{atm}/\rho)=0$ (solid), 
$-1$ (dashed-dotted), $-2$ (dashed), and $-4$ (dotted).
Heavy lines denote response-function-weighted radii
and light lines denote flux-weighted radii.
\label{f_relativesize}}
\end{figure}

\begin{figure}
\figurenum{7} 
\includegraphics[scale=0.65]{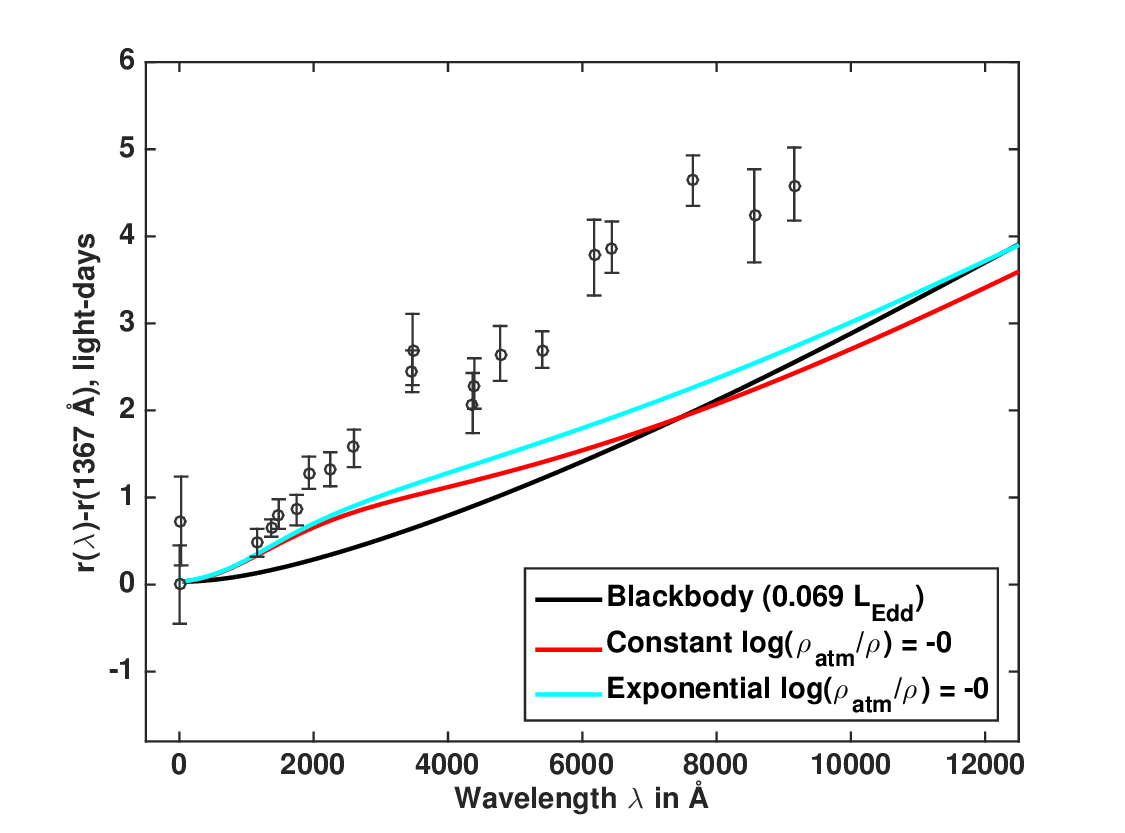}
\caption{In the same fashion as Figure 1 (see that caption for details), we plot
the {\em absolute} wavelength-dependent relative disk reverberation time lag estimated from the
response-function-weighted mean radius $r(\lambda)$ (solid lines)
for blackbody disks (black) and scattering-atmosphere disks with
constant density (red) and exponential density (cyan), all with $\log (\rho_{atm}/\rho)=0$.
Points from F16 are plotted assuming zero time delay for the {\em Swift} hard X-ray band.
Even at the density of a standard thin disk, the deviation from blackbody emission 
due to electron scattering in the radiation pressure-dominated region of the disk leads
to reverberation lags that are up to two and a half times longer than predicted in the
blackbody model (corresponding to the thick solid lines in Figure \ref{f_relativesize}.)
\label{f_real}}
\end{figure}

The models discussed here do not by themselves match all AGN observations.
They produce SEDs that peak at shorter wavelengths than observed in many AGN. 
While the models can show a non-zero lag between X-ray and ultraviolet emission
(e.g., Figure \ref{f_time} for $\log (\rho_{atm}/\rho)=-4$),
as seen in NGC 2617 \citep{2014ApJ...788...48S} 
and NGC 4151 \citep{edelson+2017},
the slow increase in the lag from ultraviolet to optical wavelengths in those objects
is not reproduced.  
A disk that transitions at small radii to a torus emitting at extreme ultraviolet wavelengths \citep[e.g.,][]{edelson+2017,gd16} may alleviate these problems.

\acknowledgments
We thank the referee and B.\,Czerny, C.\,Done, C.\,Knigge, J.\,Matthews, H.\,Netzer, and D.\,Starkey for discussions.
P.H. and G.S. acknowledge the support of the Natural Sciences and Engineering Research Council of Canada (NSERC), funding reference number 2017-05983. 
P.H. acknowledges the support and hospitality of NRC Herzberg Astronomy \& Astrophysics during his sabbatical.
K.H. acknowledges support from STFC grant ST/M001296/1.
\software{MATLAB}
\begin{soundtrack}
\music{http://www.vnvnation.com/}{VNV Nation}
\music{https://en.wikipedia.org/wiki/Music_of_Star_Wars}{John Williams' Star Wars Soundtracks}
\music[.]{http://www.dragonforce.com/}{Dragonforce}
\end{soundtrack}

\appendix

\section{Density and Temperature in Gas Pressure-dominated Regions of an Accretion Disk}

We desire equations for $\rho$ and $T$ in regions of an accretion disk where
$P=P_{gas}$ and $\kappa=\kappa_{\rm es}+\kappa_{\rm bf+ff}$.
For comparison with SS73,
we use cgs units and neglect general relativistic corrections.

We will require $D(r)$, the rate at which mechanical energy is transformed into heat energy per unit surface area over both sides of the disk. 
Following \cite{fkr}, we write
\begin{equation} \label{eqd}
D(r) = \frac{ 3G\dot{M}M }{8\pi r^3} \left( 1-\sqrt{\frac{r_{in}}{r}} \right)
= \frac{ 3G\dot{M}M }{8\pi r^3} f(r).
\end{equation}
where $M$ is the black hole mass, $\dot{M}$ is the mass accretion rate, and 
$f(r)$ 
involves the innermost disk radius $r_{in}$.

To find $\rho$, we begin with the radiative energy transport equation
$\ssb T^4= \kappa\Sigma F$ (NT73 Eq.\ 5.8.1c) where $T$ is the interior disk temperature, $\kappa=\kes+\kappa_{\rm bf+ff}$ is the opacity,
$\Sigma$ is the mass surface density,
and $F(r)$ is the total surface flux emitted at $r$.  
We take $\kappa_{\rm bf+ff}=\kf \rho T^{-7/2}$
so that $\kf$ has units of cm$^{2}$ g$^{-1}$ (g cm$^{-3}$)$^{-1}$ K$^{7/2}$.
We assume that energy released is radiated locally, so that $F(r)=D(r)$.
Substituting in for $\kappa$, $\Sigma = 2\rho h$, and $F(r)=D(r)$,
\begin{equation}
\ssb T^4= (\kes+\kf\rho T^{-7/2}) \times 2\rho h \times (3G\dot{M}Mf(r)/8\pi r^3). 
\label{eqetrans}
\end{equation}
The above is a quadratic equation for $\rho$.
We substitute for the disk half-thickness
$h=(Pr^3/\rho GM)^{1/2}$ (NT73 Eq.\ 5.8.1a)
rewritten using $P=\rho k_B T/\mu_p$ as
$h=(k_B T r^3/GM\mu_p)^{1/2}$, where $\mu_p$ is the mean mass per particle in grams.
The quadratic equation and its solution are:
\begin{eqnarray}
a_q \rho^2 + b_q \rho + c_q = 0 \label{rhoquad} {\rm ~with~} 
a_q = \kf T^{-7/2} ~;~ 
b_q = \kes  ~;~ 
c_q = -4\pi \ssb T^{7/2}\mu_p^{1/2}r^{3/2}/3(k_BGM)^{1/2}\dot{M}f(r) \\
\rho = \frac{\kes T^{7/2}}{2\kf} \left[-1 + \sqrt{1+\frac{16\pi\kf \ssb (\mu_p/k_B)^{1/2}}{3\kes^2 (GM/r^3)^{1/2}\dot{M}f(r) } }\right] \label{rhosolve1} 
\end{eqnarray}

To find $T$, we use two expressions for the vertically integrated shear stress,
$W=2h\alpha P=2h\alpha\rho k_B T/\mu_p$ (NT73 Eqs.\ 5.7.5d and 5.8.1b)
and $W=[f(r)\dot{M}(GM/r^3)^{1/2}]/2\pi$ (NT73 Eq.\ 5.6.14a),
to find a second equation relating $\rho$ and $T$:
\begin{equation} \label{rhosolve2}
\rho = \frac{f(r)\dot{M}(GM/r^3)}{4\pi\alpha (k_BT/\mu_p)^{3/2}}.
\end{equation}
Equating Eqs.\ \ref{rhosolve1} and \ref{rhosolve2} and solving for $T$, we find:
\begin{equation}
T = \left[ \frac{\kf f(r)\dot{M}(GM/r^3)}{2\pi\kes\alpha(k_B/\mu_p)^{3/2}}
\left(-1 + \sqrt{1+\frac{16\pi\kf \ssb (\mu_p/k_B)^{1/2}}{3\kes^2 (GM/r^3)^{1/2}\dot{M}f(r) }}\right)^{-1} \right]^{1/5}
\end{equation}

We have verified that these expressions have the correct numerical coefficients and dependencies on $r$, $M$, $\dot{M}$, and $\alpha$ in the limits $\kf\ll\kes$ and $\kf\gg\kes$.
The limiting solutions are approached slowly; e.g., the disk $T$ is higher at large radii (SS73 region c) than predicted by the limiting solution because $T$ has a steeper radial dependence in region b than in region c, resulting in the limiting $T$ in region c being approached from above.

In disk regions where $P=P_{rad} + P_{gas}$ and $\kappa=\kes$,
we searched for analytic solutions for $\rho$ and $T$ 
but found only implicit equations.


\listofchanges

\end{document}